\documentclass{jfm}
\usepackage{graphicx}
\usepackage{epstopdf, epsfig}

\usepackage{amstext}
\usepackage{amsmath}
\usepackage{amsfonts,amssymb}
\usepackage{framed}
\usepackage{algorithm}
\usepackage{algorithmicx}
\usepackage{algcompatible}
\usepackage{subcaption}
\usepackage{lipsum}
\usepackage{bm}
\usepackage{xcolor}
\usepackage{pdfpages}
\usepackage{lineno}
\usepackage{textcomp}

\def\dashL{\bm{\mbox{--~--~--}}}

\def\dashdot{- $\cdot$ -}

\def\Lbox{\mbox{---}~{\hspace*{-.1in}\tiny $\square$}\hspace*{-.1in}~\mbox{---}}
\def\Lcirc{\mbox{---}~{\hspace*{-.12in}$\circ$}\hspace*{-.12in}~\mbox{---}}

\def\Ltriag{\mbox{---}~{\hspace*{-.14in} \small $\triangle$}\hspace*{-.1in}~\mbox{---}}

\usepackage{multicol}

\shorttitle{Surface waves with microplastics and surfactants}
\shortauthor{Y. Sun, C.Ruf, T. Bakker and Y. Pan}

\title{Effects of microplastics and surfactants on surface roughness of water waves}

\author{Yukun Sun\aff{1},
    Christopher Ruf\aff{2},
    Thomas Bakker\aff{1},
    \and
    Yulin Pan\aff{1}
    \corresp{\email{yulinpan@umich.edu}}
    }

\affiliation{
\aff{1}Department of Naval Architecture and Marine Engineering, University of Michigan, Ann Arbor MI 48105, USA
\aff{2}Department of Climate and Space Sciences and Engineering, University of Michigan, Ann Arbor MI 48105, USA
}

\begin{document}

\maketitle

\begin{abstract}

In this paper, we study the flow physics underlying the recently developed remote sensing capability of detecting oceanic microplastics, which is based on the measurable surface roughness reduction induced by the presence of microplastics on the ocean surface. In particular, we are interested in whether this roughness reduction is caused by the microplastics as floating particles, or by the surfactants which follow similar transport paths as microplastics. For this purpose, we experimentally test the effects of floating particles and surfactants on surface roughness, quantified by the mean square slope (MSS), with waves generated by a mechanical wave maker or by wind. For microplastics, we find that their effect on wave energy and MSS critically depends on the surface area fraction of coverage, irrespective of the particle sizes in the test range. The damping by particles is observed only for fractions above $O(5-10\%)$, which is much higher than the realistic ocean condition. For surfactants, their damping effect on mechanically generated irregular waves generally increases with the concentration of surfactants, but no optimal concentration corresponding to maximum damping is observed, in contrast to previous studies based on monochromatic waves. In wind-wave experiments, the presence of surfactants suppresses the wave generation, due to the combined effects of reduced wind shear stress and increased wave damping. For the same wind speed, the wind stress is identified to depend on the concentration of surfactants with a power-law relation. The implications of these findings to remote sensing are discussed.

\end{abstract}

\begin{keywords}
Authors should not enter keywords on the manuscript, as these must be chosen by the author during the online submission process and will then be added during the typesetting process (see http://journals.cambridge.org/data/\linebreak[3]relatedlink/jfm-\linebreak[3]keywords.pdf for the full list)
\end{keywords}
 
\section{Introduction} \label{sec:intro}

Ocean plastics pollution is an urgent and global problem. An estimated eight million tons of plastics trash enters the ocean each year, and most of it is battered by sun and waves into microplastics. Information about the distribution and volume of microplastics is vital to address the removal of plastic pollution from the ocean environment. Recently the development of global observational systems for microplastics has been under active discussions among interdisciplinary communities \citep{maximenko2019toward,martinez2019measuring,king2021can}. Along this direction, a new remote sensing technique \citep{ruf2020detection,EvansRuf2021} is introduced, which enables the tracking of microplastics over time and across the globe. The principle of the technique is to infer the microplastics concentration through the ocean surface roughness anomaly (i.e., lower-than-expected roughness) induced by microplastics, which can be accounted for by the difference between the real-time measurement from a spaceborne radar and a standard model of surface roughness.

\begin{figure}
    \centering{\includegraphics[width=0.65\textwidth]{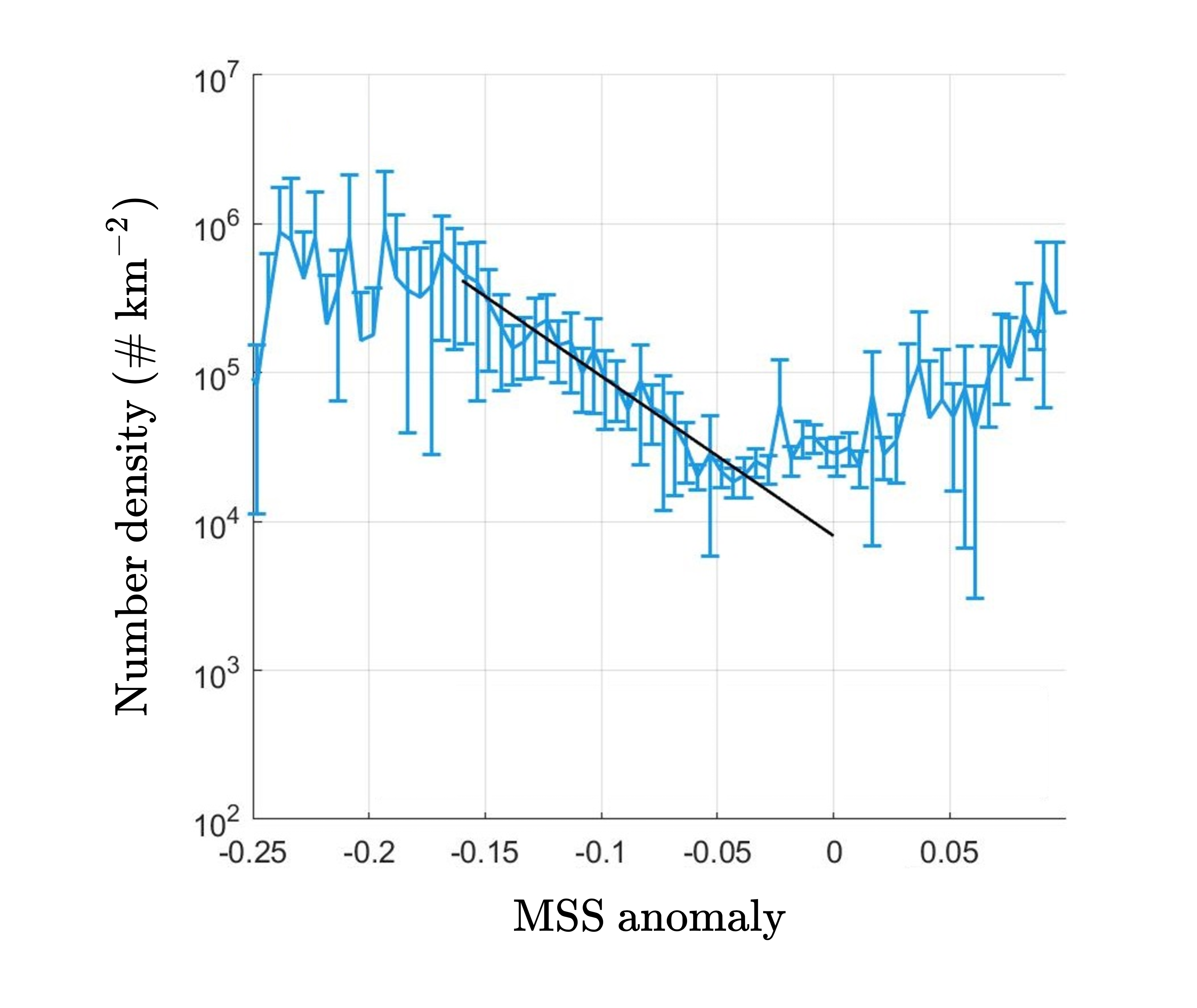}}
    \caption{Relation between the MSS anomaly (computed from CYGNSS data and Katzberg model in the oceanic region within latitudes of $\pm 38^\circ$, and averaged from June 1, 2017 to May 31, 2018) and number density of microplasitics (computed by a global microplastic transport model \citep{van_Sebille_2015}). The error bars represent $95\%$ confidence intervals with each one computed by data from the van Sebille model at locations with a given MSS anomaly. The range with a correlation between the MSS anomaly and number density is indicated ({\color{black}\rule[0.5ex]{0.5cm}{0.8pt}}). The data for this plot is obtained from \cite{EvansRuf2021}.}
    \label{fig:mss_anomalies}
\end{figure}

The new technique has been applied to data from the NASA CYGNSS (Cyclone Global Navigation Satellite System) \citep{Ruf2013, Ruf2016}. Specifically, CYGNSS measures through the GPS L1 signal the normalized bi-static radar cross section (NBRCS), whose inverse provides the mean square slope (MSS) of the ocean surface, defined as
\begin{equation}
    \textrm{MSS}=\int_0^{k_c} k^2 S(k) dk,
    \label{eq:mss}
\end{equation}
where $k$ is the wavenumber, $S(k)$ is the omni-directional spectrum, and $k_c$ is the cut-off wavenumber depending on the incident angle and carrier wave frequency in remote sensing. For CYGNSS, $k_c$ takes an average value of $7.5$ rad m$^{-1}$ \citep{mss_calibration}. As defined by \eqref{eq:mss}, $\textrm{MSS}\rightarrow \overline{\nabla\eta\cdot\nabla\eta}$ (variance of surface elevation gradient) for $k_c\rightarrow \infty$, and otherwise MSS quantifies the surface roughness up to a finite scale $k_c$. In addition to CYGNSS measurements, another source of MSS can be obtained by the standard Katzberg model \citep{katzberg2006} which takes wind speeds from a NOAA reanalysis model \citep{liu2018impact} as inputs. The MSS anomaly, defined as the relative difference between the CYGNSS measurements and Katzberg model results (normalized by the latter), is expected to account for the effect of microplastics on surface roughness, among other factors (such as error of the CYGNSS measurements and the influence of other physical processes). In \cite{EvansRuf2021}, it is found that the MSS anomaly shows a favorable correlation with the concentration of oceanic microplastics computed from a global transport model, as shown in Figure \ref{fig:mss_anomalies}.

While this new technique shows promising applications, the underlying flow physics, in particular regarding the correlation between MSS anomaly and microplastics concentration, is not clear. In \cite{ruf2020detection}, the authors suggest that the correlation may be due to the enhanced damping of surface waves by microplastics as floating particles, referring to the earlier experimental results in \cite{Sutherland2019_PRFluids}. Indeed, in a sloshing wave tank, \cite{Sutherland2019_PRFluids} show that the presence of layers of floating particles enhances the wave damping, with the enhancement increasing with the increasing number of particle layers. However, the experimental setup with the water surface fully covered by particles is not consistent with the situation of microplastics that only covers a small fraction of the ocean surface. To represent the latter situation, experiments are needed to study the characteristics of surface waves in the presence of particles partially covering the free surface. Such experiments are currently not available.

Another mechanism that can lead to the observed correlation, as hypothesized in \cite{EvansRuf2021}, is the wave damping effect by surfactants which share similar transport paths as microplastics \citep{van_Sebille_2020}. Compared to a scarce number of studies on the effect of floating particles, there is a much larger body of literature on the effect of surfactants to surface waves. It has been observed in several field studies \citep{CoxMunk54, Barger1970, huhnerfuss1981attenuation, Huhnerfuss1983_field, Ermakov1986, lombardini1989, Bock1999_experiment} that the presence of surfactants on the ocean surface results in some damping effect of surface waves. Physically, the damping effect is believed to be caused by the Marangoni stresses (due to the inhomogeneous adsorption of surfactants at the interface) that can act in opposite directions of the wave motion \citep{surfactant_hidden_variables}. In controlled experiments \citep{Dowling2001} and numerical simulations \citep{lucassen1968longitudinal,Lucassen1982EffectOS} where quantitative studies are possible, it is found that an optimal surfactant level associated with maximum damping exists for a given wave frequency. While these studies are conducted for different wave frequencies (in the range of 4 to 200 Hz) and identify different optimal surfactant levels, it is suggested by \cite{surfactant_hidden_variables} that the optimal level may correspond to the situation of the surfactant-induced Marangoni wave having the same wavelength as the surface wave, leading to some resonance-like effect. For irregular waves as in the ocean, however, the effect of varying levels of surfactant is not sufficiently studied, and it is not clear whether an optimal level for wave damping exists for a wave spectrum. This question is also relevant to the correlation problem as such an optimal level implies a non-monotonic relation between concentration and MSS anomaly that somewhat contradicts the observed correlation.

%[field study]: Cox & Munk 1954, Barger et al 1970, Huhnerfuss et al 1981, Huhnerfuss et al 1983, Ermakov et al 1986, Lombardini et al 1989, Bock et al 1999
% [wave tank]: Liu & Duncan 2006
% [simulation]: Creamer & Wright 1992, Ceniceros 2003

For wind waves, experiments in wave tanks show that the presence of surfactant significantly suppresses the wave generation \citep{mitsuyasu1986, tang1992suppression, wei_wu1992, Wu1997_circular_tank, uz2002_laboratory}. This phenomenon is usually explained through the reduction of wind shear stress by surfactants. As shown in \cite{uz2002_laboratory}, with a surfactant level of $10^{-1}$ mol l$^{-1}$, the shear stress can be reduced up to 30$\%$ compared to clean water for the same wind speed of 9.6 m s$^{-1}$. The minimum wind speed (and shear stress) to excite waves is also found to significantly increase with the presence of surfactant \citep{Wu1997_circular_tank,tang1992suppression}. With surfactant level of $9.0\times10^{-4}$ mol l$^{-1}$, \cite{mitsuyasu1986} report that the minimum wind speed leading to a wave growth is between 10 and 12.5 m s$^{-1}$. In spite of the scattered results, most studies focus on the wave characteristics at one (or only a few) surfactant levels. As a result, there are no systematic relations identified for the relevant parameters including wind speed, shear stress, wave growth rate, surfactant level and the associated surface tension. Uncovering these relations is not only helpful for interpretation/application of the aforementioned CYGNSS data, but also desirable in understanding wave generations on the ocean surface that is inevitably contaminated by surfactants. 

In the present study, we aim to understand the mechanism of the MSS anomalies detected by CYGNSS as well as the general physics of wave damping by floating particles and surfactants. For these purposes, we conduct wave-tank experiments to test the effects of floating particles and surfactants to damp the waves generated by either a mechanical wave maker or wind. For floating particles, we use two sizes of particles with 0.5 cm and 0.8 cm (both in the range of oceanic microplastics) and consider their partial coverage of the free surface with varying area fraction. We find a critical dependence of the damping effect of particles on the area fraction, irrespective of the size of particles. The enhanced particle damping of MSS and energy of the surface waves is only observed for fractions above $O(5\sim10\%)$. For low fraction of $O(0.1\%)$ that corresponds to real oceanic microplastics situations, the wave energy is not affected and the MSS is slightly increased (probably due to the diffraction by particles). For surfactants, results on mechanically generated waves show that the presence of surfactant damps the wave energy and MSS, but no optimal surfactant level (associated with maximum damping) can be identified as in the previous cases with monochromatic waves. The surfactants also suppress the wind-generated waves, leading to a higher critical wind speed to excite waves and a lower growth rate (for excited waves). Physically, we attribute this suppression to not only the previously argued effect of reduced wind shear stress, but also the wave damping by surfactants. Furthermore, at the same wind speed, we find that the wind shear stress depends exponentially on the surface tension, or depends on the concentration of surfactants in a power-law relation (with a negative exponent linearly correlated with the wind speed). The presented results clearly identify the presence of surfactant as the dominant contribution to the MSS anomaly detected by CYGNSS.   

The organization of the rest of this paper is as follows. The details of the experimental setup and procedures, including methods of measurements and choice of floating particles and surfactants, are described in \S\ref{sec:experimental setup}. The reference results with both mechanically and wind generated waves in clean water are discussed in \S\ref{sec:sample_results_clean}. The main results with the presence of microplastics and surfactants are presented in \S\ref{sec:microplasics_results} and \S\ref{sec:surfactants_results}, respectively. Finally, the conclusion is provided in \S\ref{sec:conclusion}.

\section{Experimental Setup and Procedures} 
\label{sec:experimental setup}
\subsection{Facility and Input Parameters} \label{sec:wave_tank}
\begin{figure}
  \centerline{\includegraphics[scale = 0.1]{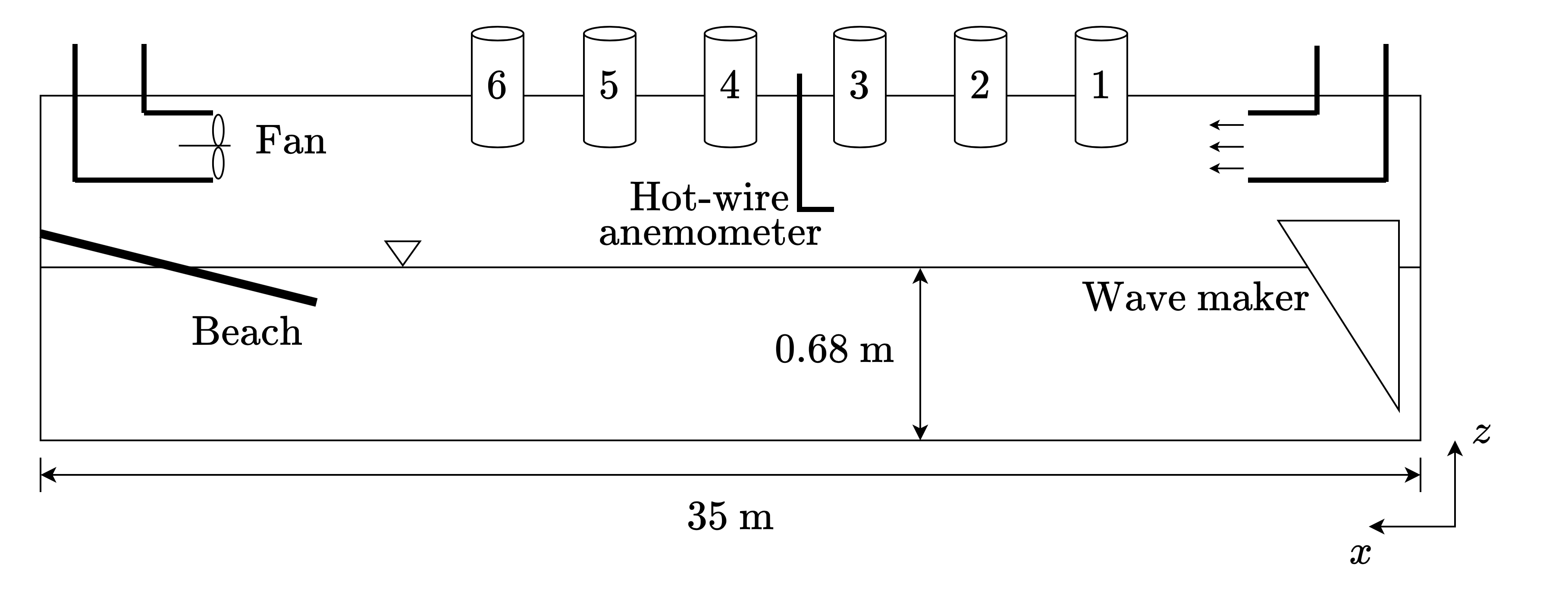}}%
  \caption{Schematic sketch of the side view of the wave tank, including a wave maker, a wind tunnel, six ultrasonic sensors, a hot-wire anemometer, and a beach. The spacings between the sensors are not perfectly scaled, with detailed information provided in Table \ref{tab:probe_distance}.}
\label{fig:WWT}
\end{figure}

The experiments are conducted in the wind-wave tank facility in the Marine Hydrodynamics Laboratory (MHL) at the University of Michigan, with a schematic sketch of the tank shown in figure \ref{fig:WWT}. The tank is 35 m long and 0.7 m wide with a water depth of 0.68 m. Waves can be generated by either a mechanical wave maker or wind through an open-loop tunnel, and are dissipated at one end of the tank by a beach with slope of $5^\circ$. 

The wave maker is wedge-shaped with an angle of $30^\circ$ to the vertical direction, and spans the width of the wave tank. The motion of the wave maker is numerically actuated by a servo motor manufactured by Kollmorgen\textsuperscript{\textregistered} AKM2G. The servo motor is feedback controlled with a proportional-integral-derivative (PID) controller, according to an input frequency spectrum $S_\textrm{in}(f)$ that describes the power spectral density of the surface elevations $\eta(t)$. For an approximation of the real ocean scenario in this study, we use the Bretschneider spectrum, a two-parameter wind-wave spectrum empirically developed for fully-developed seas: 
% \begin{equation}
%     S_{\textrm{in}}(\omega) = \frac{1.25}{4} \frac{\omega_p^4}{\omega^5} H_s^2 \mathrm{e}^{-1.25\left(\omega_p/\omega\right)^4},
%     \label{eq:Bretschneider}
% \end{equation}
\begin{equation}
    S_{\textrm{in}}(f) = \frac{1.25}{8\pi} \frac{f_p^4}{f^5} H_s^2 \mathrm{e}^{-1.25\left(f_p/f\right)^4},
    \label{eq:Bretschneider}
\end{equation}
where $f_p$ is the frequency of the peak mode and $H_s$ the significant wave height. In the current study, we use $f_p=1.25$ Hz, corresponding to $\lambda_p=1$ m as deep-water waves, and $H_s=3.9$ cm, corresponding to a moderate effective steepness $\epsilon=H_sk_p/2\approx0.12$ with $k_p\equiv 2\pi/\lambda_p$ the peak wavenumber. In each experiment, the wave maker is actuated for 200 seconds, with acceleration and deceleration for 10 seconds at the start and the end of the actuation.

\begin{figure}
  \centerline{\includegraphics[width=0.9\textwidth]{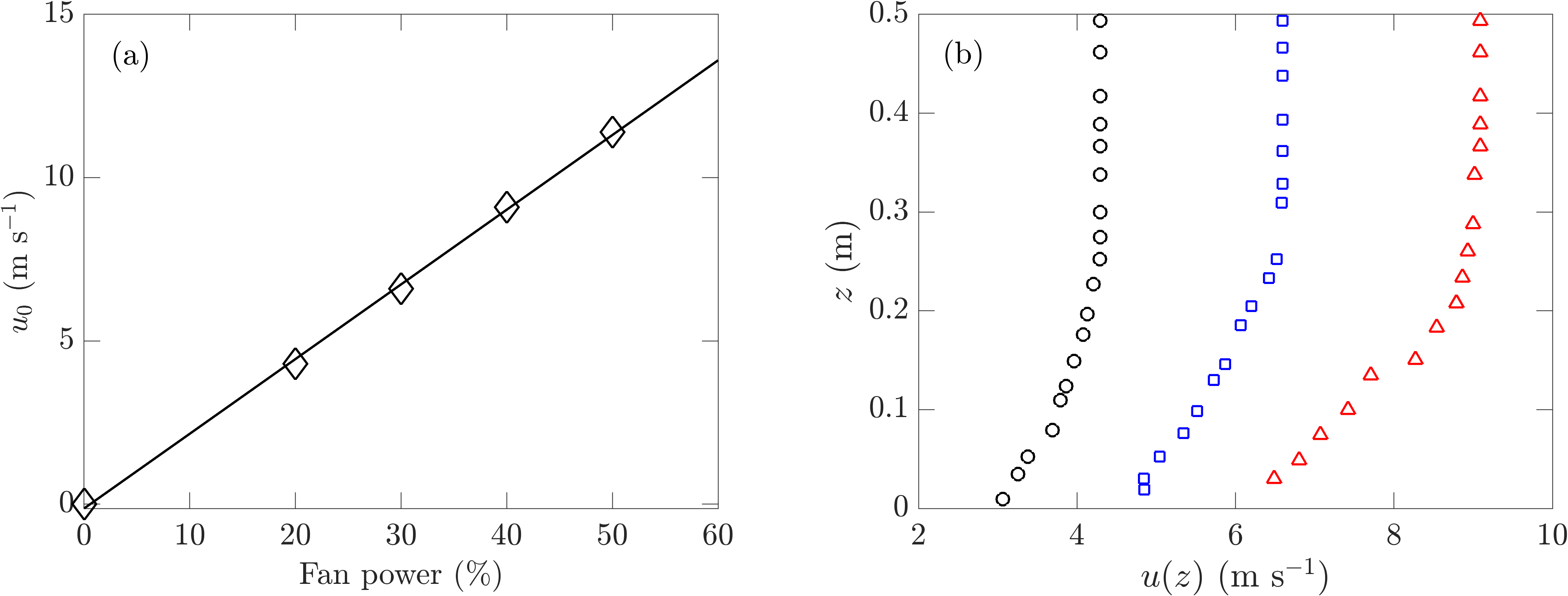}}% Images in 100% size
  \caption{(a) Reference wind speeds $u_0$ for different fan powers ({\color{black}$\diamond$}), with the linear fit ({\color{black}\rule[0.5ex]{0.5cm}{0.8pt}}). (b) Wind profiles with fan powers of $20\%$ ({\color{black}$\circ$}), $30\%$ ({\color{blue}$\square$}), and $40\%$ ({\color{red}$\triangle$}), correspondig to $u_0=4.29$, 6.59, and 9.09 m s$^{-1}$, respectively.}
\label{fig:wind_clean}
\end{figure}

Winds of different speeds are generated in an open-loop wind tunnel, which is powered by a 40-hp fan and controlled by the percentage of its maximum output power. The bottom of the wind tunnel outlet is 10 cm above the water level. The wave tank is well sealed to prevent air leakage that may cause pressure fluctuations and therefore unsteady freestream wind speeds. A hot-wire anemometer manufactured by Extech\textsuperscript{\textregistered} is placed at 13 m downstream of the wind tunnel outlet to measure the wind speed. Operated at a sampling rate of 1 Hz, the anemometer is mounted on a vertical traverse that allows one-degree-of-freedom movements in the $z-$axis. Wind profiles are measured by adjusting the location of the traverse, with all measurements taken over the stationary state of the wind. It is found that the maximum wind speed along the vertical axis (hereafter the reference wind speed) satisfies a linear relation with the fan power, as shown in Figure \ref{fig:wind_clean}(a). In the current study, we use three reference wind speeds of 4.29, 6.59, and 9.09 m s$^{-1}$ at fan powers of $20$, $30$, and $40\%$, with measured wind profiles plotted in figure \ref{fig:wind_clean}(b).

% The data acquisition system consists of 6 Senix ToughSonic\textsuperscript{\textregistered} model 14 ultrasonic sensors. The wave probes are mounted on top of the wave tank and deliver voltage signals to a National Instruments\textsuperscript{\textregistered} data acquisition board. The intervals between the sensors are non-uniform and the distance from each sensor to the wave maker as well as the wind tunnel outlet are listed in Table \ref{tab:probe_distance}. The sampling frequency of the wave probes is fixed at 100 Hz although the maximum sampling frequency is up to 2,000 Hz. The data acquisition starts along with the actuation of either the wave maker or the fan. The recording of data lasts 160 sec for mechanically generated waves and 200 sec for wind waves.

The data acquisition system consists of 6 Senix ToughSonic\textsuperscript{\textregistered} model-14 ultrasonic sensors mounted on the top of the wave tank. The intervals between the sensors are non-uniform, with distances from each sensor to the wave maker and the wind tunnel outlet listed in Table \ref{tab:probe_distance}. The measurement of surface elevation $\eta(t)$ is converted from the voltage signals delivered by the sensors to a National Instruments\textsuperscript{\textregistered} data acquisition board. The sampling frequency of the sensors is fixed at 100 Hz (which is sufficient for the current study) although the maximum sampling frequency is up to 2000 Hz. 

% For mechanically generated waves, the data acquisition starts at $t_w=30$ sec to skip the time needed for the establishment of the irregular wave field and lasts for 160 sec, where $t_w$ denotes the instance for the actuation of the wave maker. For wind waves, each run starts with calm water and the data acquisition is started along with the wind at $t_w=0$. The data acquisition in turn lasts 200 seconds for sufficient amount to time to allow wind waves to establish and to develop.

% Check name of "data acquisition board"

\begin{table}
  \begin{center}
  \def~{\hphantom{0}}
  \begin{tabular}{lcccccc}
      Sensor                       & 1 & 2 & 3 & 4 & 5 & 6 \\[3pt]
      Distance to wave maker (m)  & 9.35 & 11.38 & 14.23 & 18.66 & 22.76 & 28.45\\[3pt]
      Distance to wind outlet (m) & 2.03 & 4.06 & 6.91 & 11.34 & 15.44 & 21.13
  \end{tabular}
  \caption{Distances from the 6 wave sensors to the wave maker and to the wind outlet}
  \label{tab:probe_distance}
  \end{center}
\end{table}

\subsection{Testing Procedures} \label{sec:exp_microplastics}
\subsubsection{Experiments with microplastics}
The major materials of floating plastic debris in real ocean are polyethylene and polypropylene \citep{Cozar2014_plastics}. In this study, we use two types of particles both made of polypropylene (density $\rho_p = 0.92$ kgm$^{-3}$): one as the PolyFil PolyPellets\textsuperscript{\textregistered} microbeads with irregular shape and average diameter $D_p\approx0.5$ cm, and the other as McMaster Carr\textsuperscript{\textregistered} plastic balls with regular shape and diameter $D_p=0.8$ cm (both within the size range of the oceanic microplasitics \citep{Cozar2014_plastics}). To quantitatively test the wave damping by particles, we focus on mechanically generated waves, since the area fraction of particles (defined in \eqref{eq:particle_conc}) is very difficult to control in wind waves (due to the drift by wind). 

\begin{figure}
    \centering{\includegraphics[scale = 0.08]{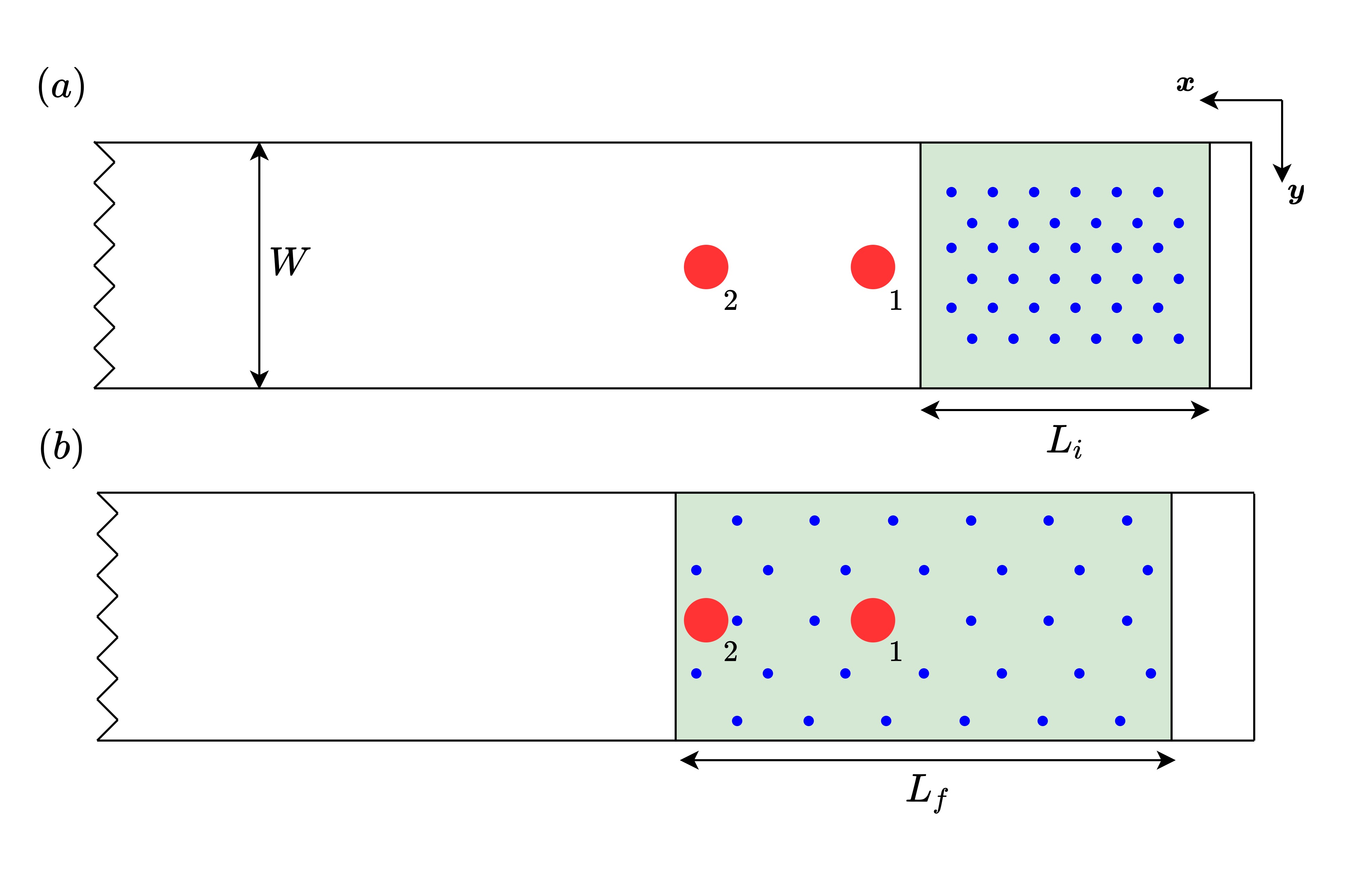}}
    \caption{Schematic sketch of location and spread of particles at the (a) beginning and (b) the end of each experiment. For simplicity, only the locations of sensors 1 and 2 represented by ({\color{red}$\bullet$}) are shown. The particle-covered regions are marked by green color with particles illustrated by ({\color{blue}$\bullet$}).}
    \label{fig:particle_conc_wwt}
\end{figure}

Each experiment with particles in mechanically generated waves is conducted with the following procedures. We first place the particles in the wave tank at 8.53 m from the wave maker in calm water. The particles then spread out both up- and downstream due to particle-particle and particle-surface interactions until a balance is reached with a recorded length of $L_i$ (see Figure \ref{fig:particle_conc_wwt}(a)). As waves pass through the particles, they further spread out and drift downstream due to the wave effect (e.g., the Stokes drift as discussed in \cite{stokes_2009,van_Sebille_2020}), with the final spreading length (after waves pass through) recorded as $L_f$ (see Figure \ref{fig:particle_conc_wwt}(b)). To quantify the concentration of particles with varying spreading length, we define an (average) area fraction
\begin{equation}
    C = \frac{N_p S_p}{W \overline{L}},
    \label{eq:particle_conc}
\end{equation}
where $N_p$ is the number of particles, $S_p=\pi D_p^2/4$ is the planform area of one particle, $W$ is the width of the tank and $\overline{L}=(L_i+L_f)/2$ is the average length of the spreading. In practice, we choose $N_p$ such that the value of $C$ ranges from 0.1$\%$ to 20$\%$, with the lowest fraction close to the oceanic microplastics concentration. A total of 9 values of $C$ in this range are tested (5 and 4 for smaller and larger particles), with each experiment repeated for 3 times to quantify the uncertainty level. 

%for two types of particles combined, with each area fraction tested after draining, flushing, and re-filling the wave tank to reduce water contamination by other impurities. 

To test the repeatability of using \eqref{eq:particle_conc} to measure the concentration, we perform 3 repetitions for each value of $N_p$ for both types of particles, with the results shown in figure \ref{fig:particle_conc_wwt} including the error bars that represent one standard deviation on each side. Along with very small error bars (which illustrates sufficient repeatability), we see a power-law relation between $C$ and $N_p$ for both types of particles and that the value of $C$ is independent of the particle sizes for large $N_p$. While these behaviors themselves are intriguing and may imply deeper physics, we leave the investigation of them to future work and focus on the effect of particles on surface waves in this study. 

Finally, due to the drift of particles, the final length $L_f$ (for larger $N_p$) may cover the surface area beneath sensors 1 and 2. In order to robustly measure the effect of particles on waves (i.e., to avoid the acoustic reflection from particles and consistently consider waves after passing through all particles), we focus on sensor measurements downstream of the particle patch $L_f$. In addition, we exclude data from sensor 6 since it is relatively far away from the particles (so that the wave properties may be modified too much by nonlinear interactions instead of particle damping). Therefore, in \S\ref{sec:microplasics_results} we present results based on sensors 3, 4 and 5 for floating particle experiments. 

\begin{figure}
  \centerline{\includegraphics[scale = 0.36]{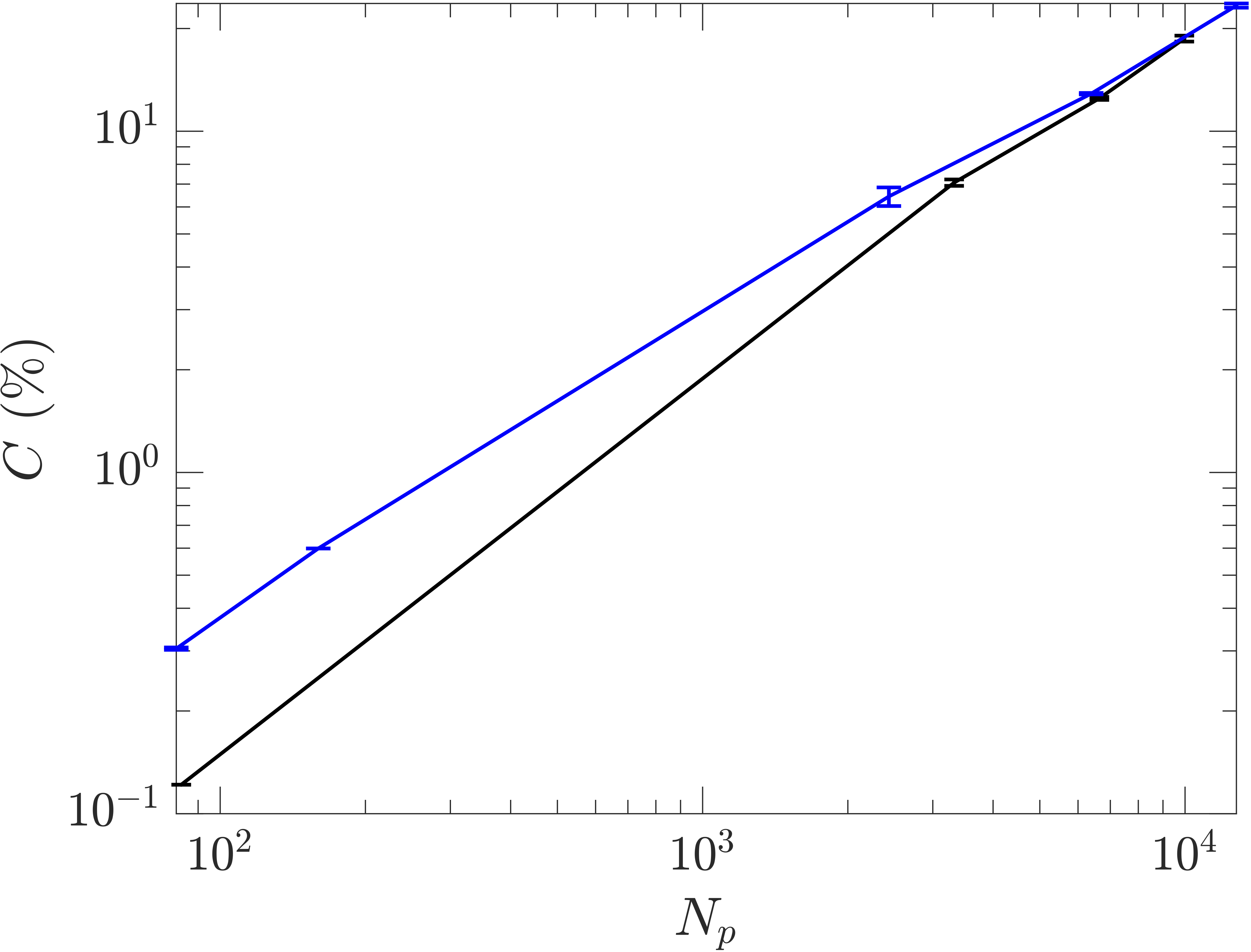}}% Images in 100% size
  \caption{Values of $C$ for particles with $D_p$=0.5 cm ({\color{black}\rule[0.5ex]{0.5cm}{0.8pt}}) and $D_p$=0.8 cm ({\color{blue}\rule[0.5ex]{0.5cm}{0.8pt}}) at different $N_p$, with error bars representing one standard deviation on each side calculated from three repetitions.}
\label{fig:particle_conc}
\end{figure}

\subsubsection{Experiments with surfactants} \label{sec:exp_surfacatnts}
\begin{figure}
  \centerline{\includegraphics[scale = 0.25]{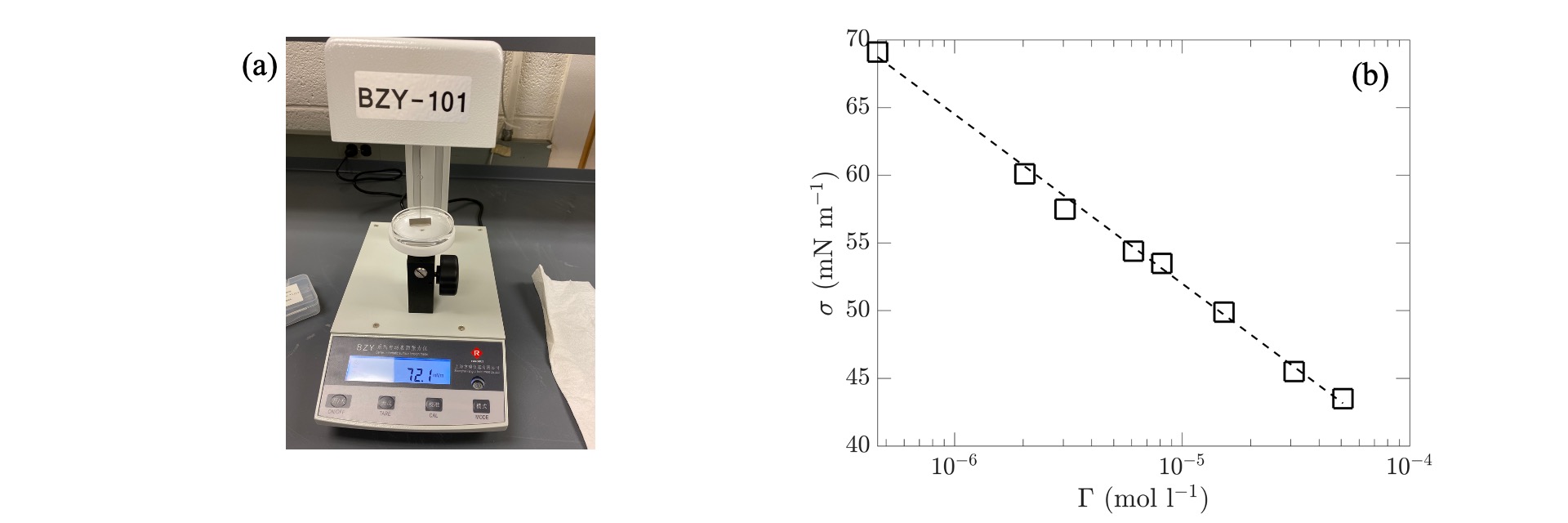}}% Images in 100% size
  \caption{(a) The surface tensiometer BZY-101. (b) The measured surface tension $\sigma$ ({\color{black}$\square$}) depending on the concentration $\Gamma$, with a logarithmic fit ({\color{black}\dashL}) to show the relation $\sigma\sim \log(\Gamma)$.}
\label{fig:surface_tension}
\end{figure}

In the current study, we use a soluble surfactant, Triton X-100 (molecular weight $625$ g mol$^{-1}$), that has been commonly used in many previous studies \cite[e.g.][]{Dowling2001,uz2002_laboratory,duncan2006JFM}. We consider nine concentrations $\Gamma = \Gamma_0 - \Gamma_8$ (in terms of mol l$^{-1}$) as listed in Table \ref{tab:surface_tension}, with $\Gamma_0$, $\Gamma_1$, $\Gamma_2$, $\Gamma_6$, $\Gamma_7$, and $\Gamma_8$ tested for mechanically-generated waves, and $\Gamma_0\sim \Gamma_7$ tested for wind waves. The surface tension $\sigma$ for each concentration is measured from water samples by an Mxbaoheng\textsuperscript{\textregistered} BZY-101 surface tensiometer (figure \ref{fig:surface_tension}(a)), which uses the Wilhelmy plate method based on the balancing of surface tension, gravitational, and buoyant forces on a platinum plate \cite[e.g.][]{Dowling1999}. The measured surface tensions (with an error within $\pm0.3$ mN/m for this device) are listed in Table \ref{tab:surface_tension} and plotted in figure \ref{fig:surface_tension}(b), showing a logarithmic relation with the value of $\Gamma$. All tested values of $\Gamma$ are below the critical micelle concentration (CMC) limit for Triton X-100, which is $\Gamma=23\times10^{-5}$ mol l$^{-1}$ corresponding to a saturated surface tension $\sigma=30.6$ mN m$^{-1}$.

In each day, we perform a total of (maximum) 12 experiments related to one concentration $\Gamma$, that include three cases with different wind speeds and one case with the wave maker, with each repeated for three times. Between each two experiments we wait for 30 minutes, which allow the surfactants to reach a uniform distribution over the tank. At the end of each day more surfactants are added to the tank until the next level of desired concentration $\Gamma$ is reached. A circulation pump is turned on overnight to well mix the surfactant and water before the starting of experiments on the next day.  

\begin{table}
  \begin{center}
  \def~{\hphantom{0}}
  \begin{tabular}{lccccccccc}
      Symbol & $\Gamma_0$ & $\Gamma_1$ & $\Gamma_2$ & $\Gamma_3$ & $\Gamma_4$ & $\Gamma_5$ & $\Gamma_6$ & $\Gamma_7$ & $\Gamma_8$ \\[3pt]
      $\Gamma\times10^{5}$ (mol l$^{-1}$) & 0 & 0.05 & 0.20 & 0.31 & 0.61 & 0.81 & 1.53 & 3.11 & 5.09 \\[3pt]
      $\sigma$ (mN m$^{-1}$)              & 72.0  & 69.1 & 60.1 & 57.5 & 54.4 & 53.5 & 49.9 & 45.5 & 43.5
  \end{tabular}
  \caption{Surfactant concentrations $\Gamma_0-\Gamma_8$ used in the experiments, with the corresponding values of surface tension $\sigma$.}
  \label{tab:surface_tension}
  \end{center}
\end{table}

Finally, we summarize all experiments conducted for microplastics and surfactants in Table \ref{tab:experiments}.
\begin{table}
    \begin{center}
        \begin{tabular}{lcccc}
        Particles/Surfactants   & Concentration     & Wave generation   & Repetitions\\[3pt]
        $D_p\approx0.5$ cm & $C=0.12$, $7.07$, $12.48$, and $18.69\%$ & Mechanical & 3 \\[3pt]
        $D_p=0.8$ cm & $C=0.30$, $0.60$, $6.44$, $12.88$, and $23.55\%$ & Mechanical & 3 \\[3pt]
        TX-100 & $\Gamma_0$ to $\Gamma_2$ and $\Gamma_6$ to $\Gamma_8$ & Mechanical & 3\\[3pt]
        TX-100 & $\Gamma_0$ to $\Gamma_7$ & Wind & 3
        \end{tabular}
        \caption{Summary of setups for experiments with microplastics and surfactants.}
        \label{tab:experiments}
    \end{center}
\end{table}

\subsection{Quantities of interest}

\begin{figure}
  \centerline{\includegraphics[scale = 0.4]{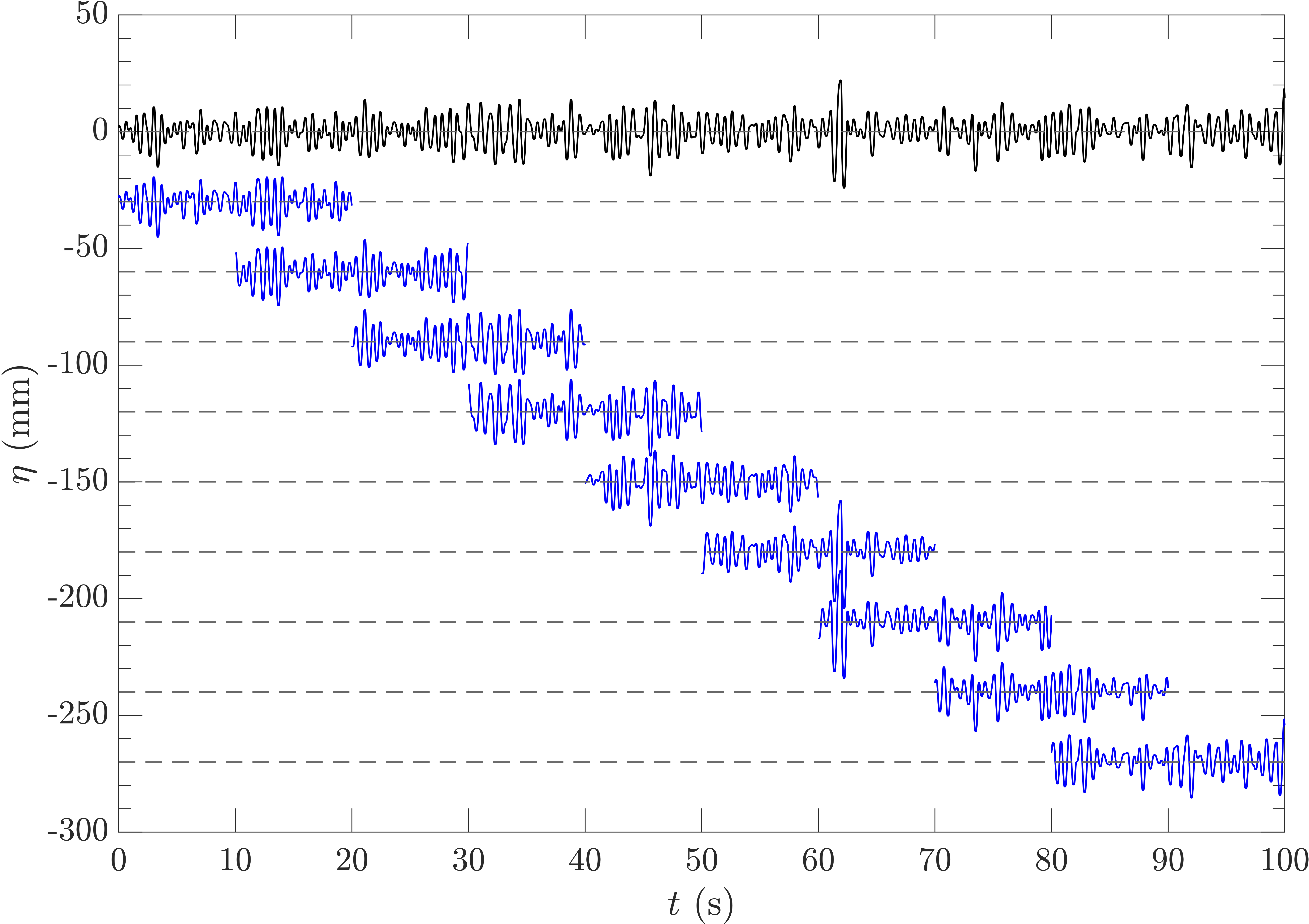}}% Images in 100% size
  \caption{A typical time series $\eta(t)$ (\color{black}\rule[0.5ex]{0.5cm}{1.0pt}) and 9 segments ({\color{blue}\rule[0.5ex]{0.5cm}{1.0pt}}), with data taken from sensor 1 in clean water test with mechanically generated waves. Each segment is vertically shifted for better visualization with the mean water level indicated by ({\color{black}\dashL}).}
\label{fig:time_series}
\end{figure}

Frequency spectra of surface elevations are calculated from the time series measured by the sensors in a time interval of 100 s of stationary state (in time). We use a standard way \citep{adak1998time} to compute the spectra as an ensemble average over spectra of 9 segments of the time series with $50\%$ overlapping of each two segments (see figure \ref{fig:time_series}). For each segment ($\Delta t=20$ s with 2000 data points), we use a Tukey window function to taper the tails of the segment, and evaluate the spectrum as 
\begin{equation}
    S_f(f) = (\Delta t/2) \left|\hat{\eta}(f)\right|^2,
\end{equation}
with $\hat{\eta}(f)$ the coefficient of cosine Fourier series of the tapered segment.

Given $S_f(f)$, we are interested in two quantities of the mean square slope (MSS) and the wave energy ($E$), defined respectively by \eqref{eq:mss} and
\begin{equation}
    E = \int_0^{k_c} S(k) \,dk,
    \label{eq:energy}
\end{equation}
with $S(k)=gS_f(f)/(8\pi^2f)$. While the evaluation by \eqref{eq:energy} is not sensitive to the cutoff wavenumber $k_c$ for $k_c$ above 7.5 rad m$^{-1}$ (correponding to CYGNSS applications), the value of MSS evaluated by \eqref{eq:mss} depends more significantly on $k_c$ up to $O(1000)$ rad m$^{-1}$. This dependence will also be discussed in \S\ref{sec:microplasics_results} and \S\ref{sec:surfactants_results} in the context of the implications of the experimental results to CYGNSS applications. 

\section{Reference results in clean water} \label{sec:sample_results_clean}
In this section, we present the reference results in clean water, with the purpose of characterizing the properties, especially quantifying the uncertainty levels, of the results.

\subsection{Wave maker experiments} \label{sec:results_clean_wedge}
\begin{figure}
    \centering{\includegraphics[width=0.9\textwidth]{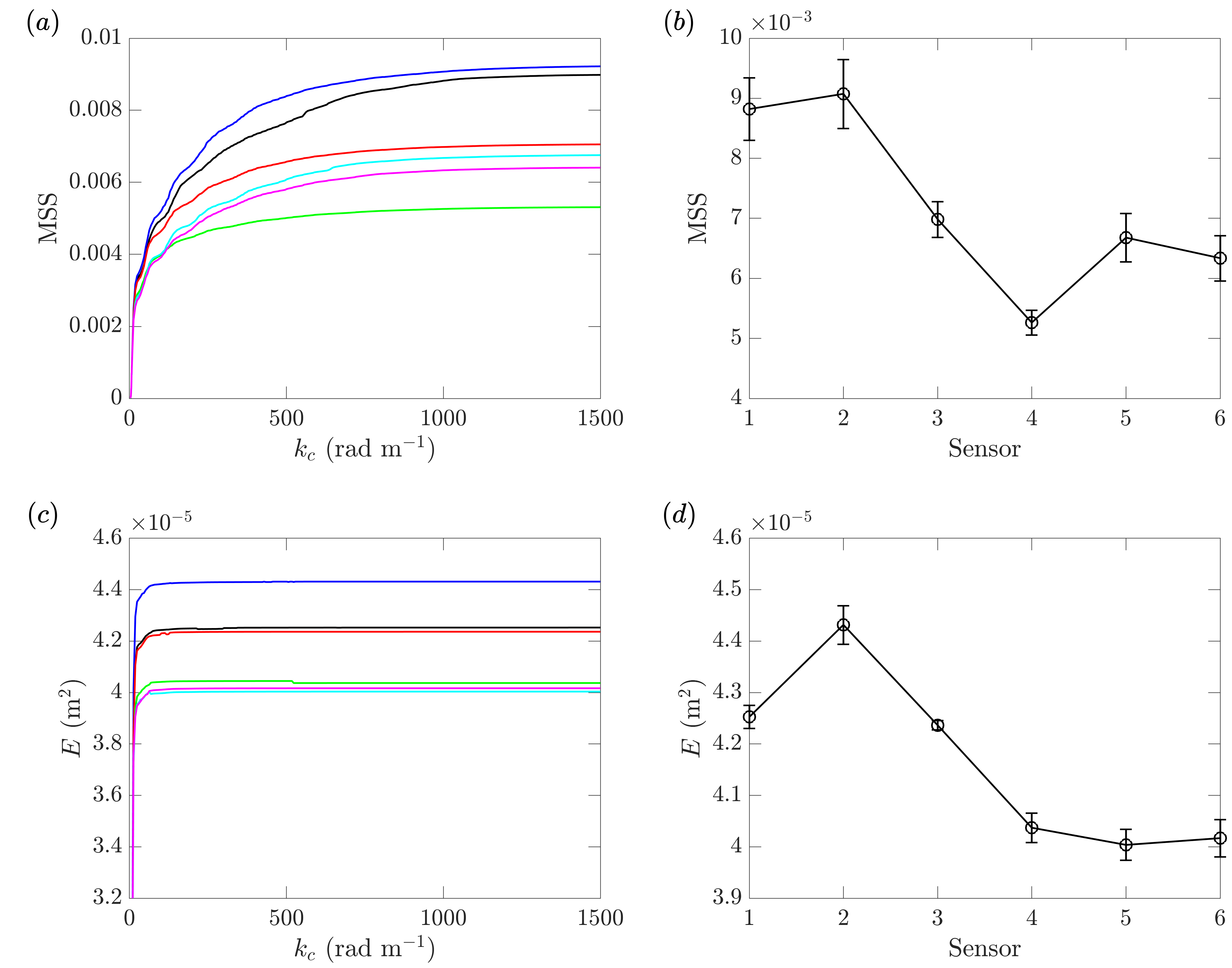}}
    \caption{(a) MSS and (c) $E$ for varying $k_c$ at sensor 1 ({\color{black}\rule[0.5ex]{0.5cm}{0.8pt}}), 2 ({\color{blue}\rule[0.5ex]{0.5cm}{0.8pt}}), 3 ({\color{red}\rule[0.5ex]{0.5cm}{0.8pt}}), 4 ({\color{green}\rule[0.5ex]{0.5cm}{0.8pt}}), 5 ({\color{cyan}\rule[0.5ex]{0.5cm}{0.8pt}}), and 6 ({\color{magenta}\rule[0.5ex]{0.5cm}{0.8pt}}) in a typical clean-water experiment. (b) MSS and (d) $E$ with $k_c=1000$ rad m$^{-1}$ at 6 sensors, including the mean values ({\color{black}\rule[0.5ex]{0.5cm}{0.8pt}}) and error bars as one standard deviation on both sides, calculated over 10 repetitions in one day.}
    \label{fig:sample_wedge_clean}
\end{figure}
We first investigate the results for mechanically generated waves, with figure \ref{fig:sample_wedge_clean}(a) and (c) showing typical results of MSS and $E$ as functions of $k_c$ measured by the six sensors in one of the runs. It can be seen that the values of MSS and $E$ converge for $k_c$ above $\mathcal{O}(10^3)$ rad m$^{-1}$ and $\mathcal{O}(10)$ rad m$^{-1}$, respectively. The main results of the paper will be presented with $k_c$=1000 rad m$^{-1}$ for both quantities (a lower value of $k_c$ results in inaccurate calculation in terms of \eqref{eq:mss} while a higher value results in higher uncertainty among repetitions of the experiments). Figure \ref{fig:sample_wedge_clean}(b) and (d) show the mean values and error bars (as one standard deviation on both sides) of MSS and $E$ with $k_c$=1000 rad m$^{-1}$, computed from 10 repetitions of each experiment within one day. For this value of $k_c$, the errors of MSS and $E$ are both acceptable (with this $k_c$ for MSS reflecting a balance between the accuracy of \eqref{eq:mss} and uncertainties in repetitions), although the uncertainty levels of MSS are generally larger especially for sensor 2 in this case. This larger uncertainty of MSS is mainly caused by the noise in the high-frequency motion of the wave maker (that is not precisely controlled due to the relatively low-frequency implementation of the feedback control), which is amplified due to the $k^2$ dependence in the MSS calculation \eqref{eq:mss}. We note that both MSS and $E$ can increase toward downstream over two adjacent sensors (say MSS for sensors 4 and 5, $E$ for sensors 1 and 2). These are due to the nonlinear wave effect which may transfer energy to high wavenumbers (leading to an increased MSS) or exchange energy among linear and nonlinear parts (leading to an increased $E$ as the linear part of the energy).

We also remark that the MSS measured over different days has a higher uncertainty level compared to that shown in \ref{fig:sample_wedge_clean}(b) (which is quantified within one day). This is probably due to the different levels of impurities (e.g., organic materials) that are flushed into the tank during the cleaning procedure at the start of each day (which is difficult to avoid for tank of this size), as well as the background noise (e.g., the wind-wave tank is located nearby a 109.7 m long towing tank which is under different operations each day). These factors may have larger influence to the higher wavenumber component of the spectrum resulting in higher uncertainty in MSS. In order to avoid such uncertainty introduced by cross-day experiments, we compare our results with and without microplastics (with multiple repetitions for each) within the same day to quantify the effect of microplastics, as will be discussed in \S\ref{sec:microplasics_results}. 

\subsection{Wind wave experiments} \label{sec:results_clean_wind}
\begin{figure}
    \centering{\includegraphics[scale = 0.36]{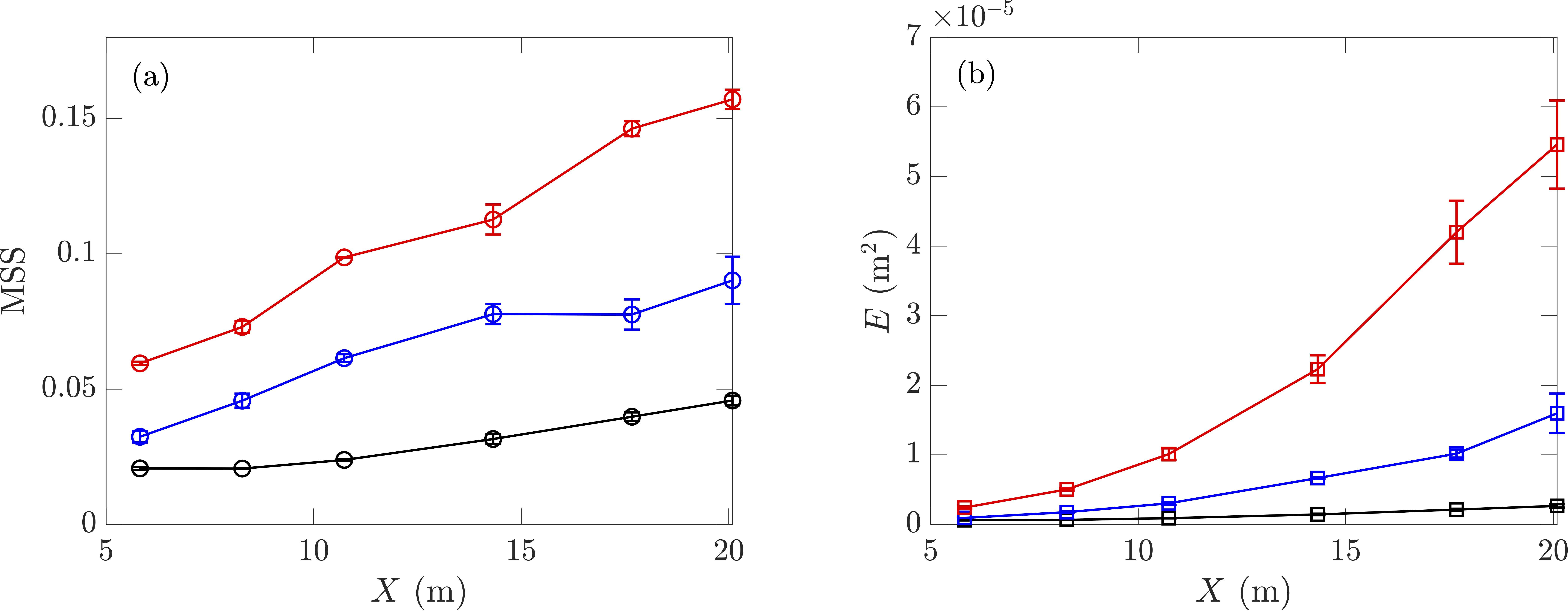}}
    \caption{(a) MSS and (b) $E$ with $k_c=1000$ rad m$^{-1}$ at 6 sensors, generated by reference wind speeds of 4.29 ({\color{black}\rule[0.5ex]{0.5cm}{0.8pt}}), 6.59 ({\color{blue}\rule[0.5ex]{0.5cm}{0.8pt}}), and 9.09 ({\color{red}\rule[0.5ex]{0.5cm}{0.8pt}}) m s$^{-1}$, including the mean values and error bars as one standard deviation on both sides, calculated over 10 repetitions in one day.}
    \label{fig:sample_wind_probe_clean}
\end{figure}

The MSS and energy $E$ measured in wind wave experiments at three different reference wind speeds are shown in figure \ref{fig:sample_wind_probe_clean}(a) and (b), as functions of the fetch (measured by the locations of sensors 1-6). The error bars are reasonably small for both quantities, indicating the robustness in the systems of wind-wave generation and measurement. For energy $E$, the uncertainty is higher than the cases with mechanically generated waves mainly because of the variation close to the wind-generated peak mode located at the relatively low frequency range of the spectrum (so it mainly adds uncertainty to $E$ but not MSS). It is also clear that both MSS and $E$ increase with the fetch and wind speeds, consistent with previous results \citep{CoxMunk54,charnock1955wind,hay1955some,ellison1956atmospheric,phillips_1958}. The condition of fully-developed sea cannot be achieved in the current 35-m tank for the three operational wind speeds.

\section{Results of experiments with floating particles} \label{sec:microplasics_results} 

\begin{figure}
    \centering{\includegraphics[width=0.9\textwidth]{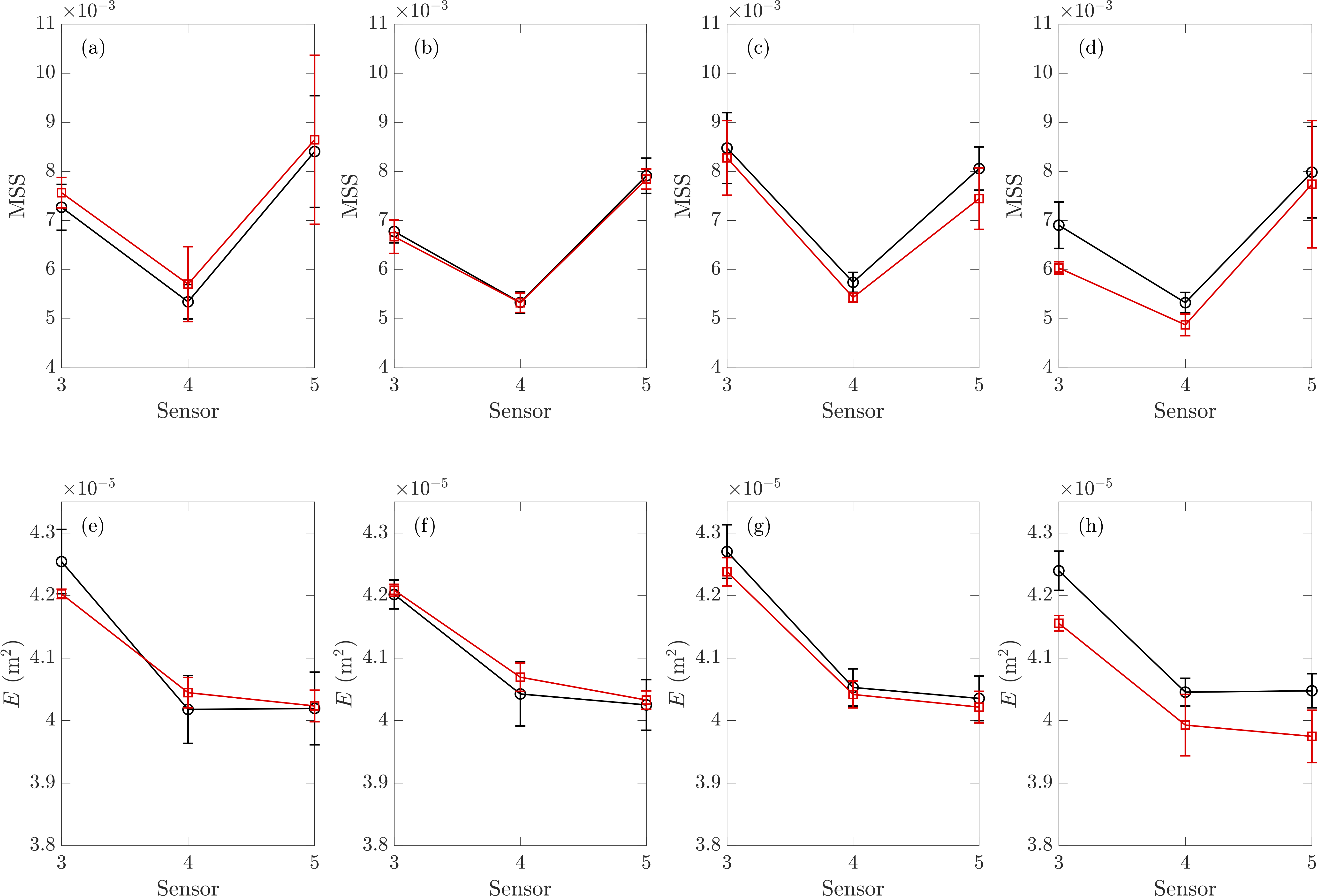}}
    \caption{The MSS ({\color{red}\Lbox}, top row) and energy $E$ ({\color{red}\Lbox}, bottom row) measured in experiments with particles of size $D_p\approx0.5$ cm by sensors 3 to 5, with area fractions of (a)(e) $0.12\%$, (b)(f) $7.07\%$, (c)(g) $12.48\%$, and (d)(h) $18.69\%$, together with the clean-water results ({\color{black}\Lcirc}) (obtained in the same day as particle experiments) for reference. The error bars of all results correspond to one standard deviations on both sides of the mean values.} 
    \label{fig:wedge_probe_micro_small}
\end{figure}

\begin{figure}
    \centering{\includegraphics[width=0.9\textwidth]{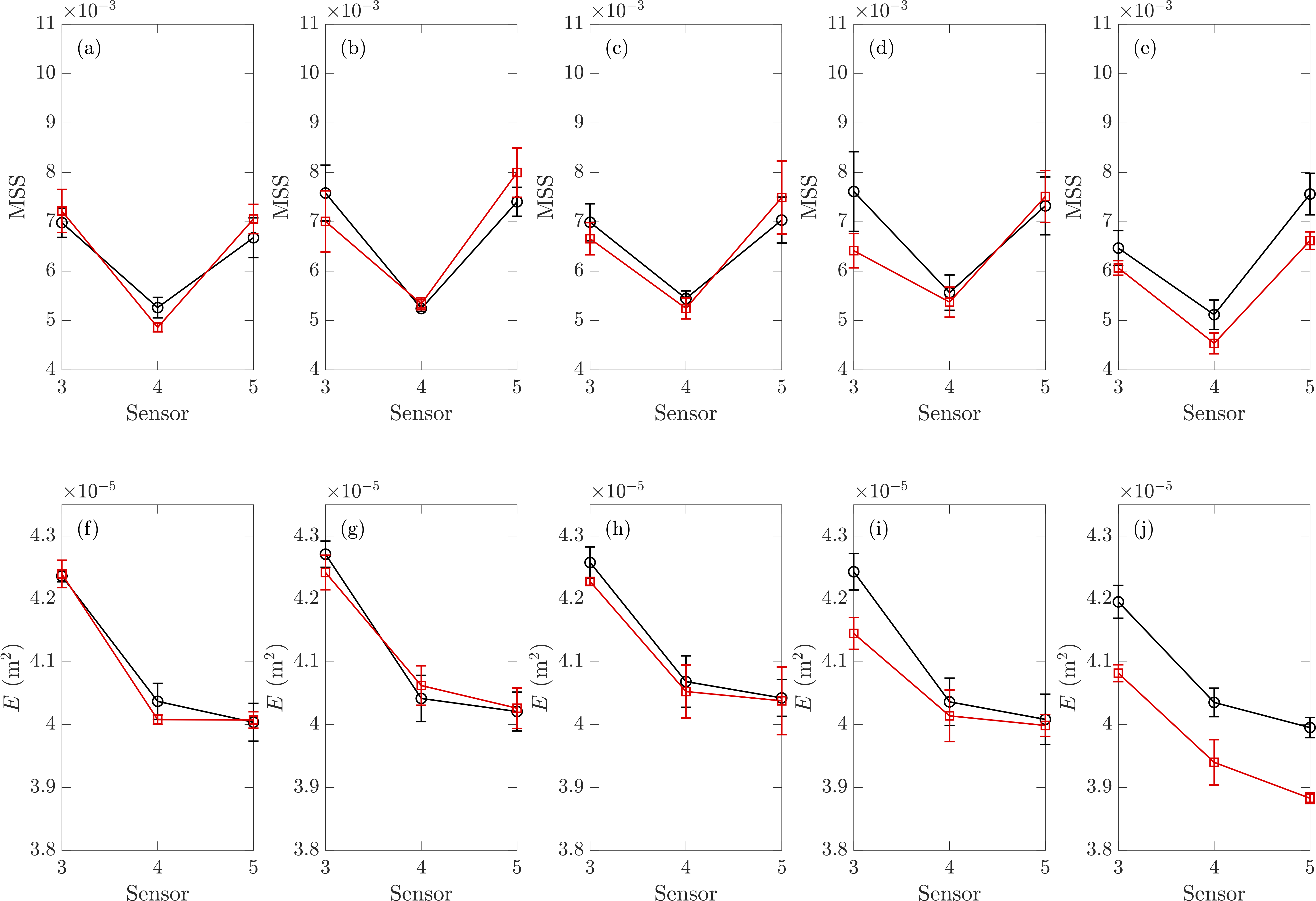}}
    \caption{The MSS ({\color{red}\Lbox}, top row) and energy $E$ ({\color{red}\Lbox}, bottom row) measured in experiments with particles of size $D_p=0.8$ cm by sensors 3 to 5, with area fractions of (a)(f) $0.30\%$, (b)(g) $0.60\%$, (c)(h) $6.44\%$, and (d)(i) $12.88\%$, and (e)(j) $23.35\%$, together with the clean-water results ({\color{black}\Lcirc}) (obtained in the same day as particle experiments) for reference. The error bars of all results correspond to one standard deviations on both sides of the mean values.}
    \label{fig:wedge_probe_micro_large}
\end{figure}

The values of MSS and $E$ at sensors 3 to 5 with small ($D_p\approx0.5$ cm) and large ($D_p=0.8$ cm) particles are  presented respectively in figures \ref{fig:wedge_probe_micro_small} and \ref{fig:wedge_probe_micro_large}, along with the clean-water results. The mean values and error bars in each sub-figure are computed from 3 repetitions of experiments with particles (with corresponding size and area fraction $C$), as well as 10 repetitions of clean-water experiments within the same day for reference (or comparison). In general, we find that the damping effects for MSS and $E$ only become obvious for sufficiently large value of $C$ for both particle sizes, e.g., in figures \ref{fig:wedge_probe_micro_small}(d), \ref{fig:wedge_probe_micro_small}(h), \ref{fig:wedge_probe_micro_large}(e), and \ref{fig:wedge_probe_micro_large}(j), where we see smaller values of the two quantities in experiments with particles relative to the clean-water experiments. For small values of $C$, the MSS and energy $E$ are barely affected for either particle size, in terms of the average values over the three sensors compared to the clean-water results (cf. figures \ref{fig:wedge_probe_micro_small}(b), \ref{fig:wedge_probe_micro_small}(f), \ref{fig:wedge_probe_micro_large}(b), and \ref{fig:wedge_probe_micro_large}(g)). The only exception to this general behavior is the MSS at small area fraction $C$ especially for the small particle (figure \ref{fig:wedge_probe_micro_small}(a)), which shows higher values at all sensors compared to the clean-water results. This is probably caused by the diffraction of short waves by the particles (or a patch of particles) that tends to increase the MSS especially when the particles are smaller. 

\begin{figure}
    \centering{\includegraphics[width=0.9\textwidth]{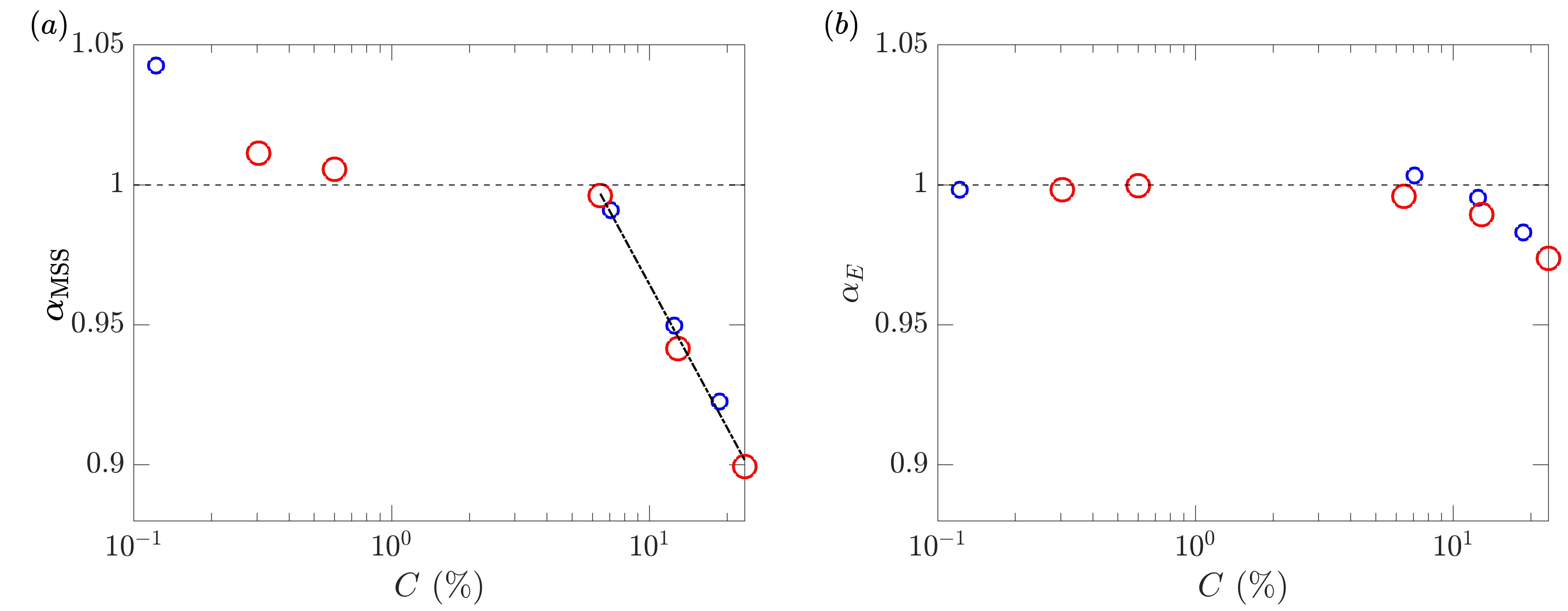}}
    \caption{(a) $\alpha_{\textrm{MSS}}$ and (b) $\alpha_E$ as functions of area fraction $C$ for particle sizes $D_p\approx0.5$ cm ({\color{blue}$\circ$}) and $D_p=0.8$ cm ({\color{red}$\circ$}). The reference value of $\alpha_{\textrm{MSS}}=1$ and $\alpha_E=1$ (i.e., no effect from particles) are indicated ({\color{black}\dashL}). A logarithmic fit to $\alpha_{\textrm{MSS}}$ at large values of $C$ is shown ({\color{black}\dashdot}).}
    \label{fig:ratio_micro_avg3probes}
\end{figure}

Since the results measured by a single sensor may be subject to higher uncertainty and influenced by other physical processes (e.g., different nonlinear effects introduced after the modification of spectra by the presence of particles) in addition to the particle damping, we further quantify the effects of particles by examining the average quantities: 
\begin{equation}
    \overline{\textrm{MSS}}_{ij} = \frac{1}{j-i+1} \sum_{n=i}^j \textrm{MSS}_n,
    \quad
    \overline{E}_{ij} = \frac{1}{j-i+1} \sum_{n=i}^j E_n,
\end{equation}
where $\textrm{MSS}_n$ and $E_n$ are MSS and $E$ measured at sensor $n$. For results in this section, we use $i=3$ and $j=5$ to obtain average values of $\overline{\textrm{MSS}}_{35}$ and $\overline{E}_{35}$. We further define the ratio $\alpha$ to quantify the effect of particles relative to the clean-water results, as
\begin{equation}
    \alpha_\phi = \overline{\phi}_{35}^p/\overline{\phi}_{35}^{c},
\end{equation}
where $\phi$ represents MSS or $E$, and the superscripts $p$ and $c$ denote results with the presence of particles and in clean water respectively. The quantities $\alpha_{\textrm{MSS}}$ and $\alpha_E$ are presented in figure \ref{fig:ratio_micro_avg3probes} as functions of the area fraction $C$ for both particle sizes. We observe that, when measured by the area fraction, the results obtained with the two particle sizes almost collapse to a single curve, indicating that the effects of particles on surface waves are irrespective of the particle size in the test range. The damping effects for both quantities become evident for values of $C$ above $C^*\sim O(5-10\%)$. The effect of particles on MSS is much larger than their effect on $E$ (e.g., at the highest value of $C$, $\alpha_{\textrm{MSS}}=0.90$ but $\alpha_E=0.97$), suggesting that the damping effect of the tested particles focuses more on the short waves. Finally, for higher values of $C$, the damping effect for MSS seems to exhibit a logarithmic relation with $\alpha_{\textrm{MSS}}-1 \sim -\log (C/C^*)$ over half a decade indicated by the data in figure \ref{fig:ratio_micro_avg3probes}(a). 

\begin{figure}
  \centerline{\includegraphics[width=0.9\textwidth]{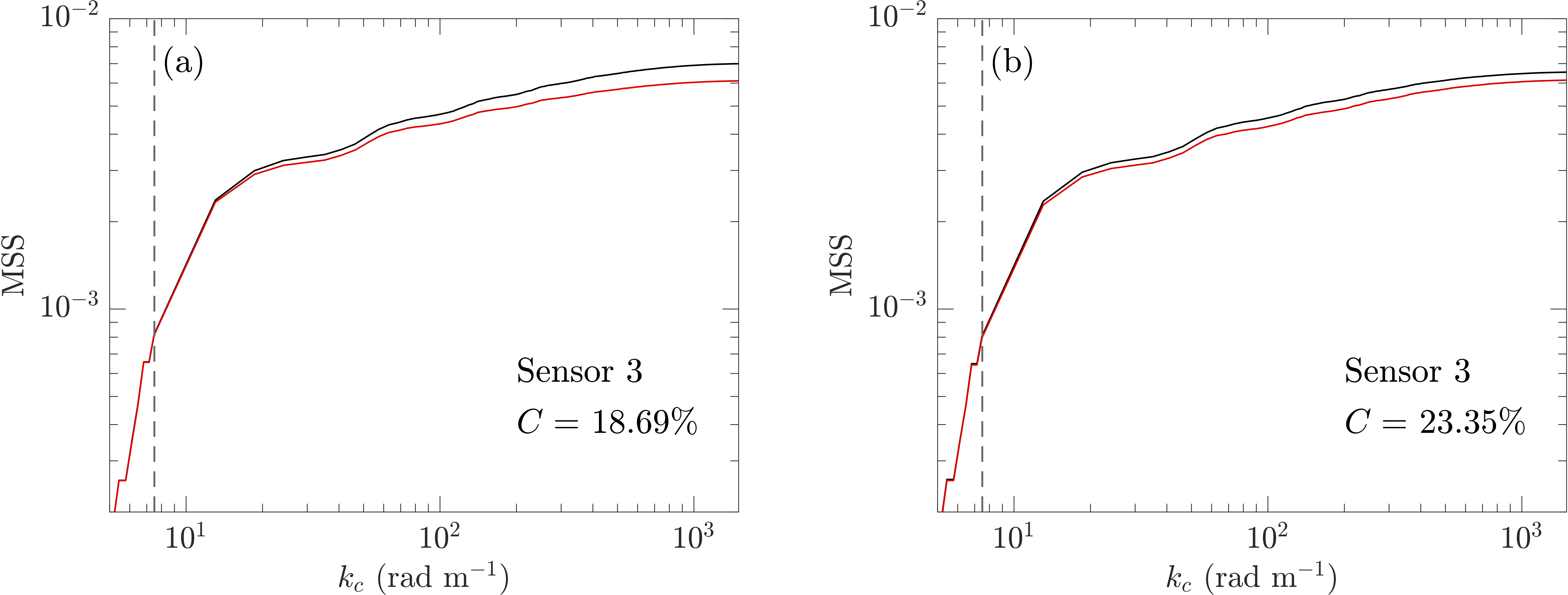}}
  \caption{MSS computed with different $k_c$ with (a) particles of $D_p\approx0.5$ cm, $C=18.69\%$ ({\color{red}\rule[0.5ex]{0.5cm}{0.8pt}}) and (b) particles of $D_p=0.8$ cm, $C=23.35\%$ ({\color{red}\rule[0.5ex]{0.5cm}{0.8pt}}). The reference clean-water results ({\color{black}\rule[0.5ex]{0.5cm}{0.8pt}}), as well as the indication of $k_c=7.5$ rad m$^{-1}$ ({\color{black}\dashL}), are shown in both (a) and (b).}
\label{fig:mss_kc_micro}
\end{figure}

Before concluding the section on floating particles, we briefly discuss the implication of the results to the remote sensing (e.g., CYGNSS) applications. Figure \ref{fig:mss_kc_micro} shows a typical result of MSS as a function of $k_c$ measured by sensor 3 for both sizes of particles at their highest area fraction $C$, along with the reference results in clean water. It is clear from the plots that for $k_c=7.5$ rad m$^{-1}$ (corresponding to CYGNSS application as discussed in \S\ref{sec:intro}), the effect of particles on MSS is negligible even for the highest concentration (let alone the oceanic concentration which is much lower). Therefore, the results are sufficient for us to conclude that the MSS anomalies observed by CYGNSS are not caused by the effects of microplastics as floating particles. 


\section{Results of experiments with surfactants} \label{sec:surfactants_results}

\subsection{Results for mechanically generated waves} \label{sec:surfactants_wedge}
\begin{figure}
    \centering{\includegraphics[width=0.9\textwidth]{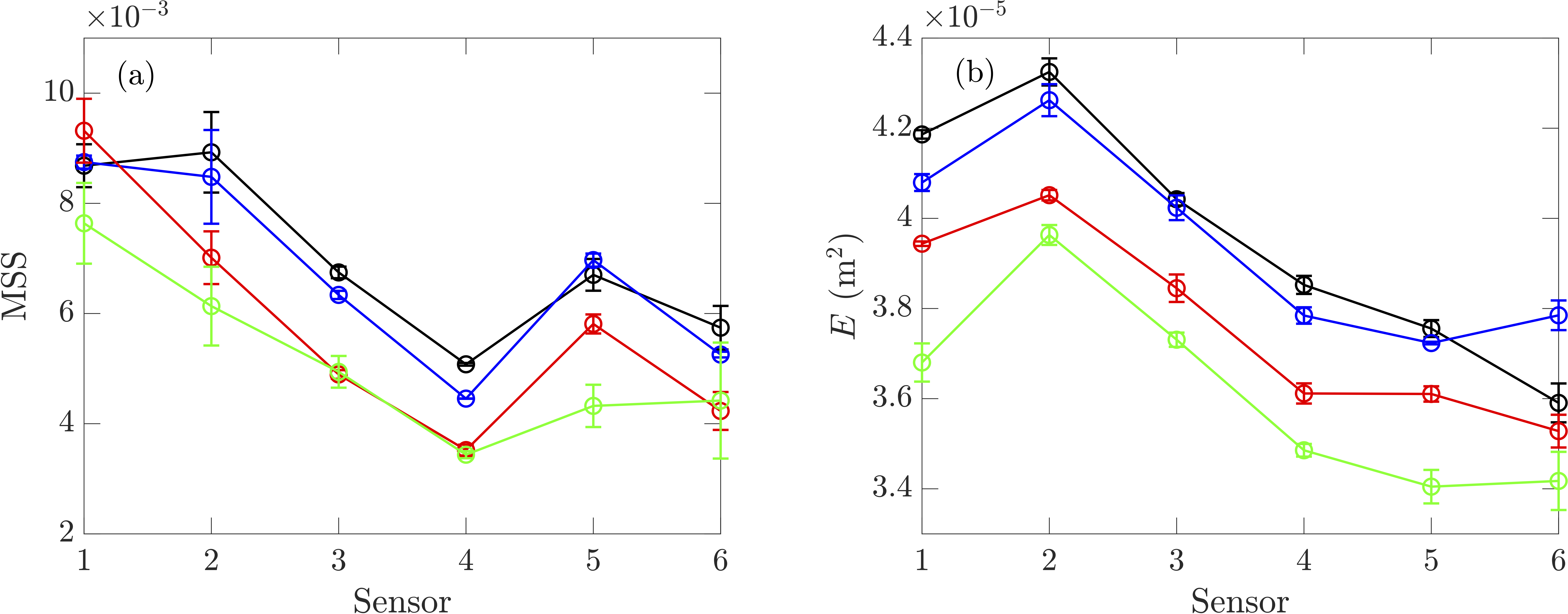}}
    \caption{(a) MSS and (b) $E$ measured by different sensors in experiments of mechanically generated waves with $\Gamma=\Gamma_0$ and $\sigma=72.0$ mN m$^{-1}$ ({\color{black}\rule[0.5ex]{0.5cm}{0.8pt}}), $\Gamma=\Gamma_1$ and $\sigma=69.1$ mN m$^{-1}$ ({\color{blue}\rule[0.5ex]{0.5cm}{0.8pt}}), $\Gamma=\Gamma_7$ and $\sigma=45.5$ mN m$^{-1}$ ({\color{red}\rule[0.5ex]{0.5cm}{0.8pt}}), and $\Gamma=\Gamma_8$ and $\sigma=43.5$ mN m$^{-1}$ ({\color{green}\rule[0.5ex]{0.5cm}{0.8pt}}). The error bars represent one standard deviation on each side of the mean value computed from 3 repetitions.}
    \label{fig:wedge_prorbe_surft}
\end{figure}

Figure \ref{fig:wedge_prorbe_surft} shows values of MSS and $E$ measured by all six sensors at different surfactant concentrations $\Gamma$ (leading to different surface tensions $\sigma$). We observe significant damping of both MSS and $E$ with the presence of surfactants, compared to the results in clean water (i.e., $\Gamma=\Gamma_0$ and $\sigma=72.0$ mN m$^{-1}$). Another interesting phenomenon seen in figure \ref{fig:wedge_prorbe_surft} is that the major surfactant-induced damping effect of MSS and energy $E$ occur in a short distance from the wave maker, namely before sensor 2 (15.47 m) for MSS and before sensor 1 (13.00 m) for $E$. After this short distance, the surfactant-induced damping effect becomes insignificant, i.e., the values of MSS and $E$ do not further deviate from the results in clean water. This phenomenon suggests a hypothetical mechanism that for each surfactant concentration, there might exist a critical level of wave amplitude below which there is no enhanced damping by surfactants. This mechanism, although yet to be verified, may be related to the Marangoni wave due to the compression/expansion of the surface, that is only strong enough with sufficiently high surface waves to result in a significant Marangoni damping.

We are also interested in the dependence of surfactant-induced damping on the concentration levels. For this purpose, we plot $\overline{\textrm{MSS}}_{16}$ and $\overline{E}_{16}$ as functions of the concentration $\Gamma$. These results are remarkably different from the previous experiments/simulations based on a monochromatic wave (with frequencies both larger and smaller than the peak frequency in \eqref{eq:Bretschneider}) 
\citep{lucassen1968longitudinal,Lucassen1982EffectOS,Dowling2001}, where an optimal concentration level exists corresponding to the maximum reduction rate for each wave frequency. Nevertheless, for irregular waves as in our study, the relation between the damping effect and the concentration level is not monotonic, with $\overline{\textrm{MSS}}_{16}$ showing a local maximum at $\Gamma=\Gamma_7$ and $\overline{E}_{16}$ at $\Gamma=\Gamma_2$. These complicated behaviors indicate that to further quantitatively understand the enhanced damping effect of surfactant to irregular ocean waves, one needs to consider the Marangoni damping in the presence of nonlinear wave interactions among a range of wave frequencies. 

\begin{figure}
  \centerline{\includegraphics[scale = 0.36]{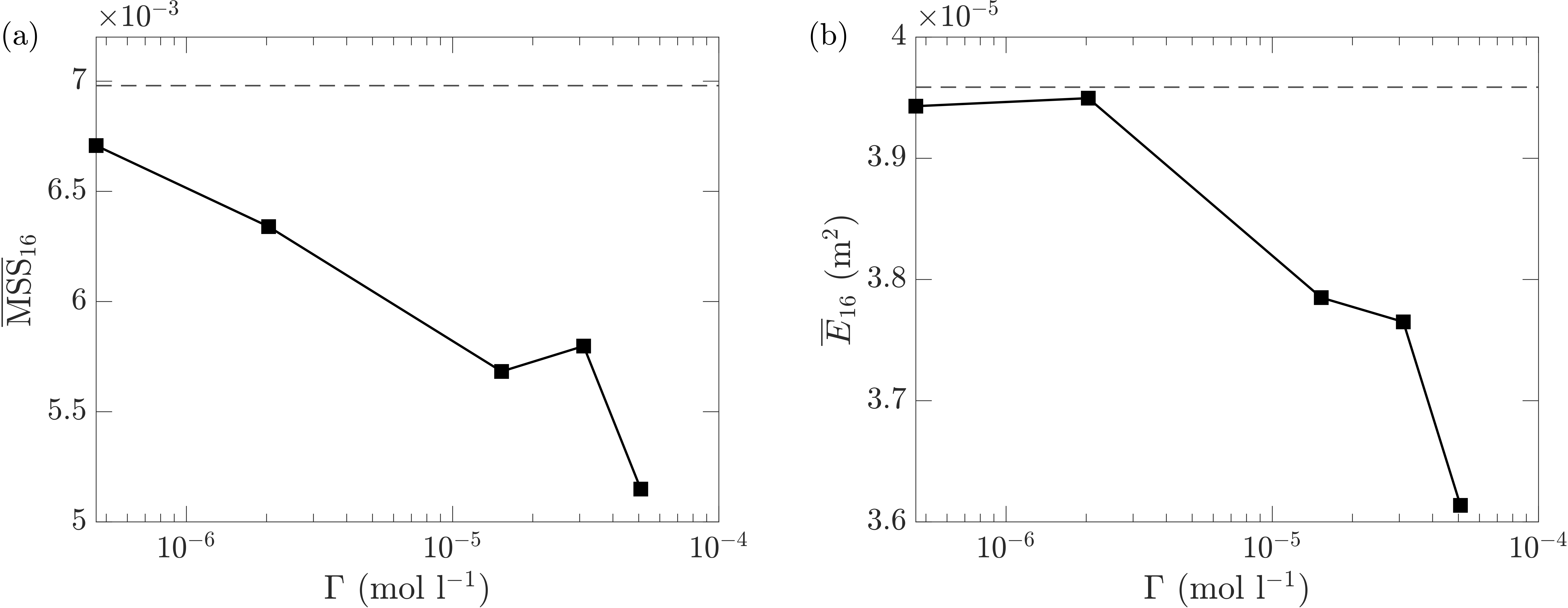}}
  \caption{(a) $\overline{\textrm{MSS}}_{16}$ and (b) $\overline{E}_{16}$ at different concentration levels of the surfactants. The results for $\Gamma=\Gamma_0$ (i.e., clean water) are indicated by ({\color{black}\dashL}) in both sub-figures.}
\label{fig:avg_growth_rate_wedge_wave}
\end{figure}

\begin{figure}
  \centerline{\includegraphics[width=0.9\textwidth]{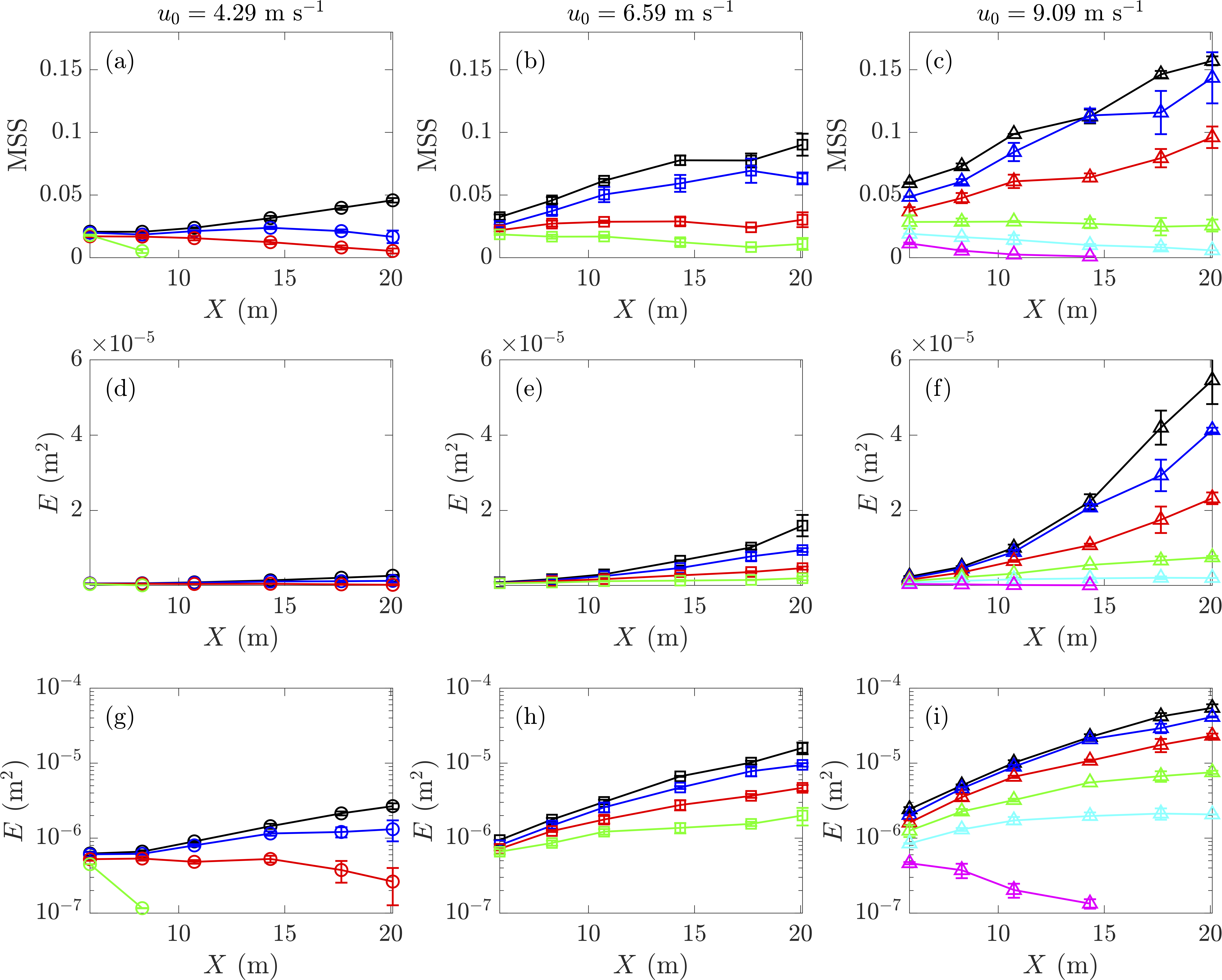}}
  \caption{MSS and $E$ with $\Gamma=\Gamma_0$ and $\sigma=72.0$ mN m$^{-1}$ ({\color{black}\Lcirc}), $\Gamma=\Gamma_1$ and $\sigma=69.1$ mN m$^{-1}$ ({\color{blue}\Lcirc}), $\Gamma=\Gamma_2$ and $\sigma=60.1$ mN m$^{-1}$ ({\color{red}\Lcirc}), $\Gamma=\Gamma_3$ and $\sigma=57.5$ mN m$^{-1}$ ({\color{green}\Lcirc}), $\Gamma=\Gamma_4$ and $\sigma=54.4$ mN m$^{-1}$ ({\color{cyan}\Lcirc}), and $\Gamma=\Gamma_5$ and $\sigma=53.5$ mN m$^{-1}$ ({\color{magenta}\Lcirc}) at fetches $X$ corresponding to sensors 1 to 6. The top and middle rows plot the MSS and $E$ respectively in a linear-linear scale, and the bottom row plots $E$ in a linear-logarithmic scale. The columns from left to right are for reference wind speeds of 4.29, 6.59, and 9.09 m s$^{-1}$, respectively.}
\label{fig:wind_probe_surft}
\end{figure}

\subsection{Results for wind waves} \label{sec:surfactants_wind}
Figure \ref{fig:wind_probe_surft} plots the MSS and energy $E$ as functions of the fetch $X$ for different wind speeds and different concentration levels of surfactant. We first note that for cases with sufficiently low wind speed and/or high surfactant concentration, there exist sensor measurements at a similar signal level to the measurements in calm water. These data points (with a criterion of $E$ less than 1.5 times that of noises measured in calm water) are physically considered as cases of no wave generation and are practically excluded in figure  \ref{fig:wind_probe_surft}. It is also noted that in figure \ref{fig:wind_probe_surft} there are some curves which decrease with the increase of $X$ as well as ``broken'' lines (say in (a), (c) and (g)), indicating wave growth at shorter fetch but then decay at longer fetch. This phenomenon is due to the convection of surfactant downstream by wind, which results in a higher concentration downstream associated with stronger damping effect. 

\begin{figure}
  \centerline{\includegraphics[width=0.9\textwidth]{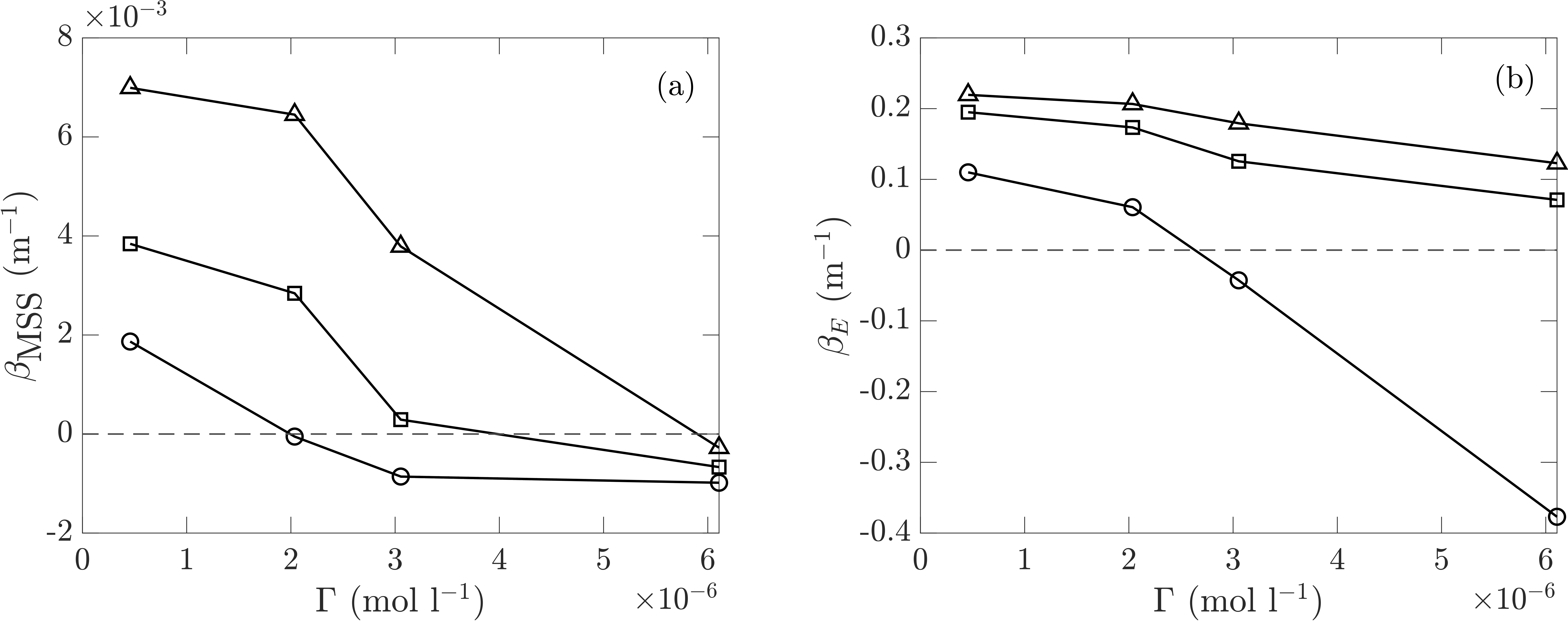}}
  \caption{(a) Linear growth rate $\beta_{\textrm{MSS}}$ for MSS (with MSS $\sim \beta_{\textrm{MSS}}X$) and (b) exponential growth rate $\beta_E$ for $E$ (with $E \sim \exp(\beta_E X)$) for varying surfactant concentrations $\Gamma$ at reference wind speeds $u_0=4.29$ ({\color{black}\Lcirc}), $6.59$ ({\color{black}\Lbox}), and $9.09$ m s $^{-1}$ ({\color{black}\Ltriag}). Only cases with wave excitation are included, and negative values indicate wave decay from sensor 1 to 6. The levels of $\beta_{\textrm{MSS}}=0$ and $\beta_E=0$ are marked ({\color{black}\dashL}).}
\label{fig:growth_rate_wind}
\end{figure}

From figure \ref{fig:wind_probe_surft} we find that MSS and energy $E$ show respectively a linear growth (cf. (a)-(c)) and an exponential growth (cf. (d)-(i)) with fetch $X$,  especially for the initial stage (i.e., short fetch) far before the fully-developed condition (i.e., dissipation much weaker than growth). While the exponential growth of energy $E$ is consistent with previous results \citep{mitsuyasu1986,lin2008direct,Deike2021_windwave}, we are not aware of any previous finding regarding the linear growth of MSS. These results indicate that in a wind sea, the peak mode grows much faster than the high-wavenumber portion of the spectrum in the initial stage of development. The linear growth rate $\beta_{\textrm{MSS}}$ for MSS (with MSS $\sim \beta_{\textrm{MSS}}X$) and exponential growth rate $\beta_E$ for $E$ (with $E \sim \exp(\beta_E X)$) are fitted from data and summarized in figure \ref{fig:growth_rate_wind}, both as functions of $\Gamma$ for each reference wind speed (only cases with wave excitation are shown). We observe that higher surfactant concentrations consistently reduce the growth rates $\beta_{\textrm{MSS}}$ and $\beta_E$. 

\begin{figure}
  \centerline{\includegraphics[scale = 0.36]{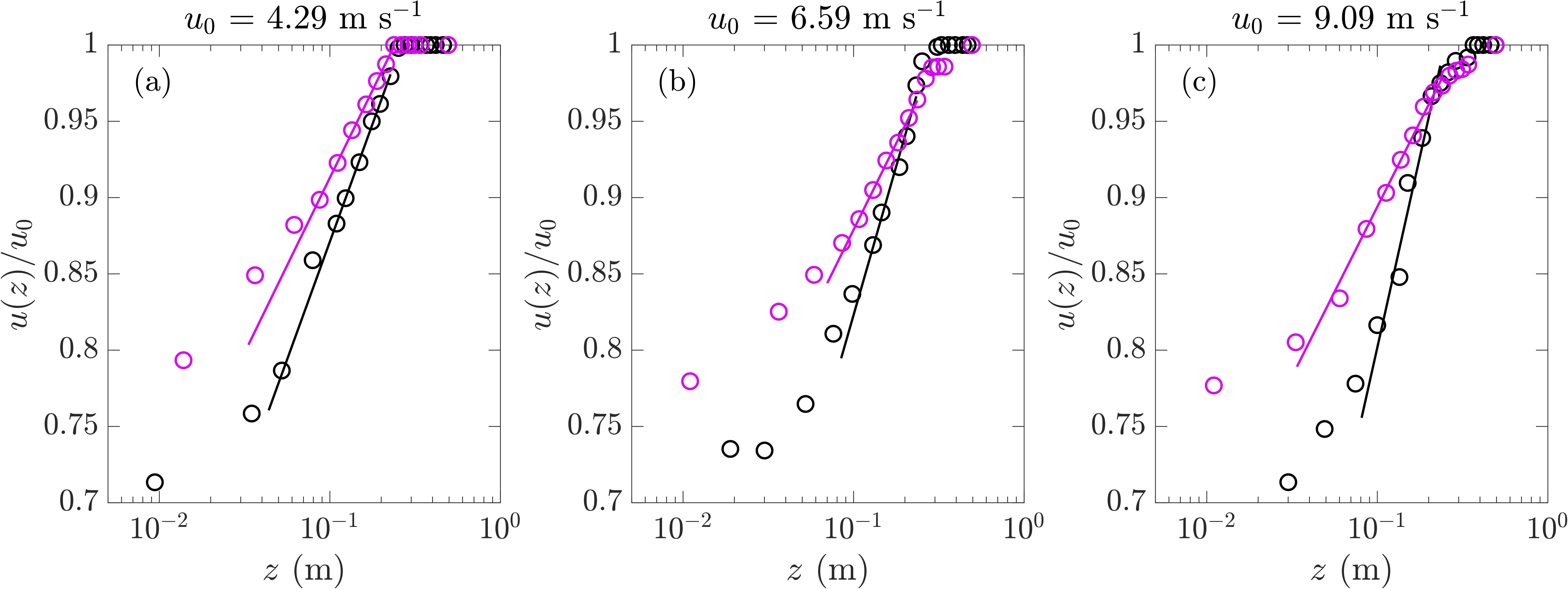}}
  \caption{Wind profiles $u(z)/u_0$ with surfactant concentrations $\Gamma_0$ ({\color{black}$\circ$}) and $\Gamma_4$ ({\color{magenta}$\circ$}), with linear fittings ({\color{black}\rule[0.5ex]{0.5cm}{0.8pt}}) and ({\color{magenta}\rule[0.5ex]{0.5cm}{0.8pt}}), at reference wind speeds (a) $u_0=4.29$, (b) 6.59, and (c) 9.09 m s$^{-1}$.}
\label{fig:wind_profile_surft}
\end{figure}

To understand the mechanisms underlying the effect of surfactants on MSS and $E$, we first investigate the wind shear stress under different levels of concentration. For this purpose, we follow the procedure of \cite{liu2016wind} to extract the friction velocity $u_*$ (and thus shear stress $\tau=\rho u_*^2$) from the logarithmic layer of the wind profile, defined as
\begin{equation}
    u(z) = \frac{u_*}{\kappa} \ln{\frac{z}{z_0}},
    \label{eq:log law}
\end{equation}
where $\kappa=0.41$ is the von K\'arm\'an constant and $z_0$ is a roughness parameter (characterizing the viscous length scale or thickness of viscous sub-layer). Figure \ref{fig:wind_profile_surft} shows some typical wind profiles as well as the fitting using $u(z)=A\ln{z} + B$ from which the friction velocity $u_*=\kappa A$ and wind shear stress $\tau$ can be calculated. 

\begin{figure}
    \centerline{\includegraphics[width=0.99\textwidth]{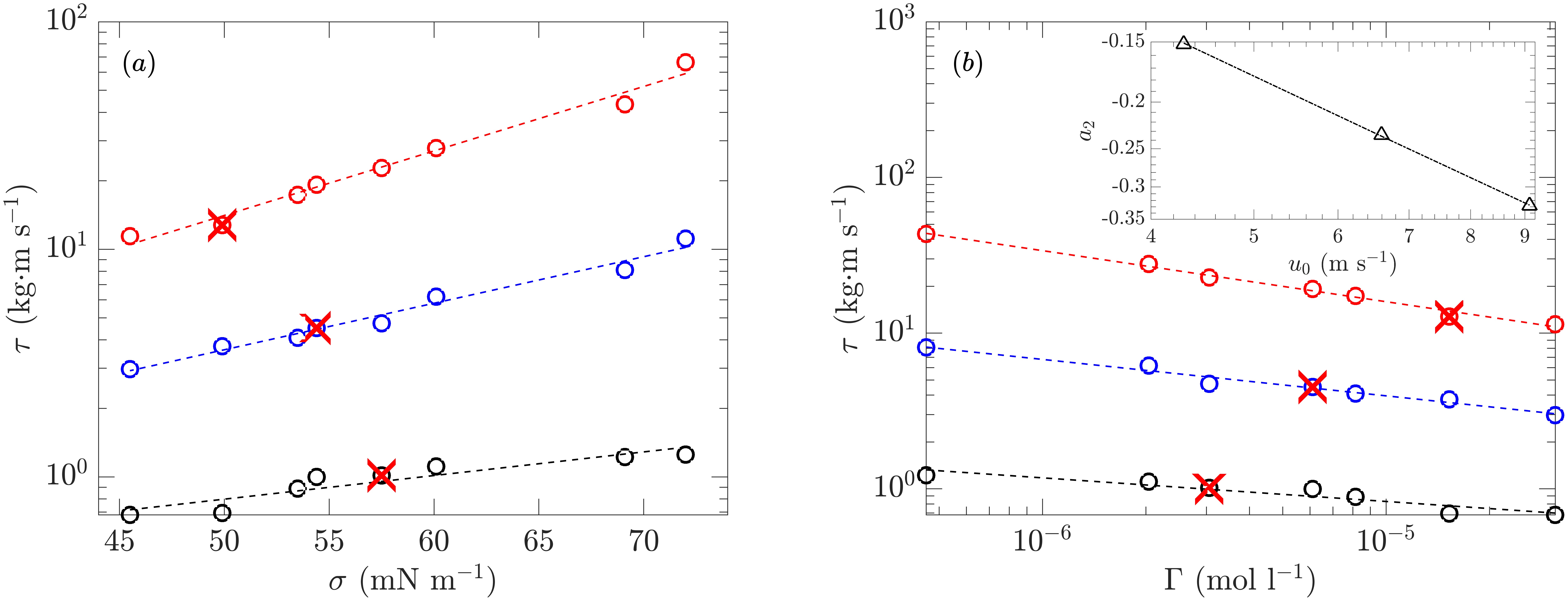}}
    \caption{Wind shear stress $\tau$ as a function of (a) $\sigma$ and (b) $\Gamma$ at $u_0=4.29$ ({\color{black}$\circ$}), 6.59 ({\color{blue}$\circ$}), and 9.09 m s$^{-1}$ ({\color{red}$\circ$}), with exponential and power-law fits marked by ({\color{black}\dashL}) for each $u_0$. Critical points of $\sigma_c$, $\Gamma_c$, $\tau_c$ corresponding to threshold of wave excitation are marked by {\color{red}$\times$} for each $u_0$. Values of $a_2$ ({\color{black}$\triangle$}) at different $u_0$ with linear fit ({\color{black}\dashdot}) are shown in the inset of sub-figure (b).}
    \label{fig:wind_stress_a2}
\end{figure}

The wind (shear) stress $\tau$ for varying surfactant concentrations (thus varying surface tensions) is plotted in figure \ref{fig:wind_stress_a2} for all three reference wind speeds $u_0$. For each wind speed we mark the critical point ($\Gamma_c, \sigma_c, \tau_c$) corresponding to the boundary below which no waves are excited. It is clear that for cases both above and below the critical points, the wind stress shows an exponential dependence on surface tension ($\tau\sim \exp(a_1\sigma)$) and a power-law dependence (which can be expected from figure \ref{fig:surface_tension}(b)) on the surfactant concentration level ($\tau\sim \Gamma^{-a_2}$) at each same wind speed. The inset of figure \ref{fig:wind_stress_a2}(b) plots the values of $a_2$ for different reference wind speeds $u_0$, which suggests a linear relation of $a_2 \sim u_0$ implying also $a_1 \sim u_0$ (although more data at other wind speeds may be needed to further consolidate this relation).

In addition, the critical points ($\Gamma_c, \sigma_c, \tau_c$) in figure \ref{fig:wind_stress_a2} reveal interesting physics on wind-wave generation. While higher wind stress implies stronger tendency for wave generation, it is possible, as suggested by figure \ref{fig:wind_stress_a2}, that waves are excited at lower $\tau$ (say $u_0=4.29$ m s$^{-1}$ and $\sigma$ around 70 mN m$^{-1}$) but suppressed at higher $\tau$ (say $u_0=6.59$ m s$^{-1}$ and $\sigma$ around 50 mN m$^{-1}$). This fact is only possible if the latter is subject to higher damping effect (which overcomes the higher wind stress) due to higher concentration of surfactants. Therefore, to understand wave generation in the presence of surfactants, it is important to consider the surfactant-induced damping as an additional physical factor. For cases in clean water with a monochromatic wave profile (for which damping can be clearly quantified), it has been demonstrated by numerical simulations \citep{Deike2021_windwave} that the wave growth rate uniquely depends on wind stress above some balanced value (which characterizes the critical state of growth from wind stress balancing the damping). The situation for our case, however, is significantly more complicated, as the damping effect depends on both the surfactant concentration level and wind speed (which affect the wavelength of the peak mode to be excited that in turn changes the damping by surfactants). These dependence relations need to be further investigated in order to fully understand the wind-wave generation in the presence of surfactants.

\begin{figure}
    \centering{\includegraphics[width=0.99\textwidth]{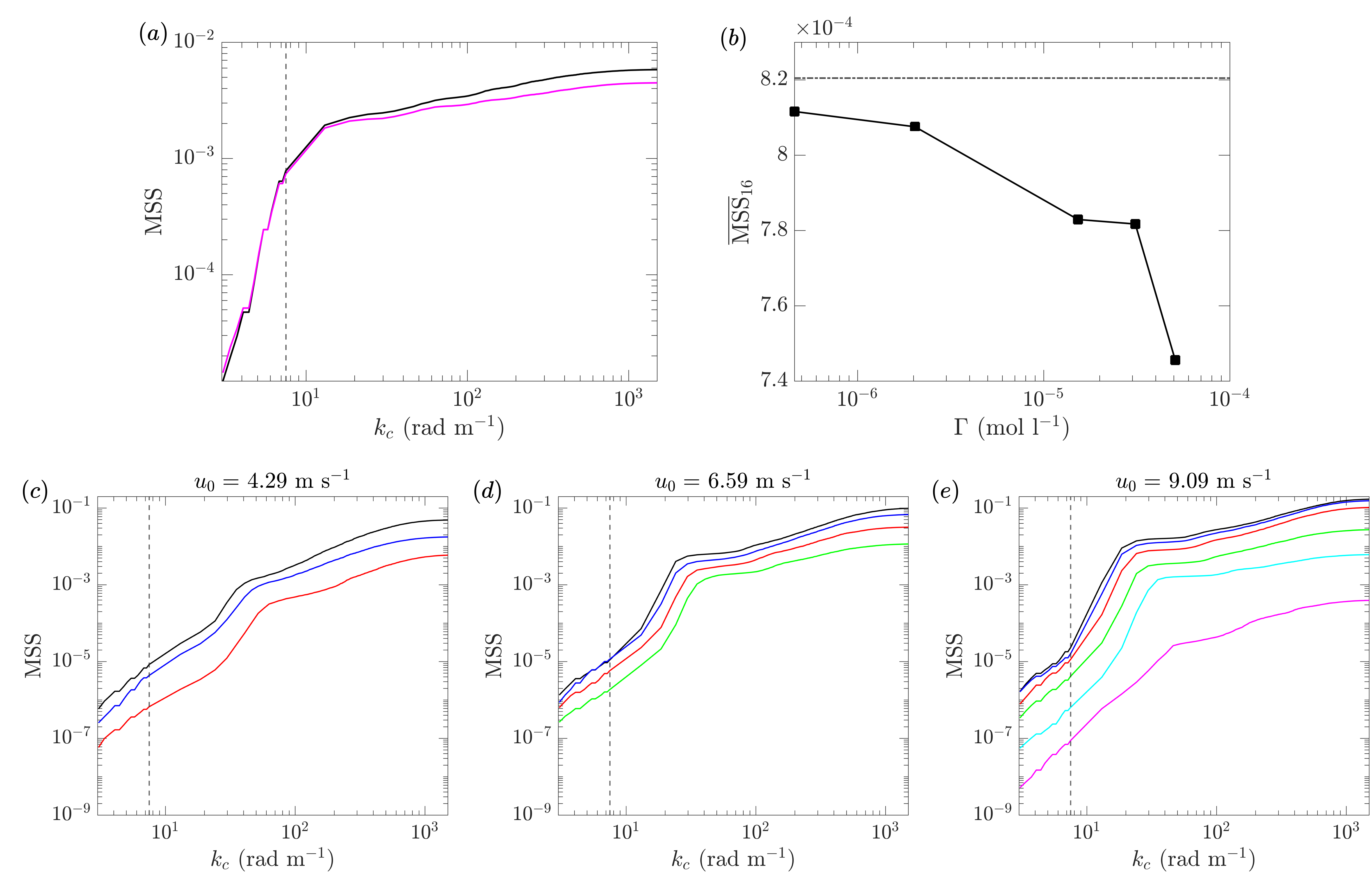}}
    \caption{(a) MSS as a function of cut-off wavenumber $k_c$ for  mechanically generated waves with surfactant concentrations $\Gamma_0$ ({\color{black}\rule[0.5ex]{0.5cm}{0.8pt}}) and $\Gamma_8$ ({\color{magenta}\rule[0.5ex]{0.5cm}{0.8pt}}); (b) $\overline{\textrm{MSS}}_{16}$ computed with $k_c=7.5$ rad m$^{-1}$ for different concentration levels $\Gamma$ for mechanically generated waves, with result from $\Gamma_0$ marked ({\color{black}\dashdot}); (c-e) MSS as a function of $k_c$ at reference wind speeds $u_0=4.29$, 6.59, and 9.09 m s$^{-1}$ respectively, with surfactant concentrations $\Gamma_0$ ({\color{black}\rule[0.5ex]{0.5cm}{0.8pt}}), $\Gamma_1$ ({\color{blue}\rule[0.5ex]{0.5cm}{0.8pt}}), $\Gamma_2$ ({\color{red}\rule[0.5ex]{0.5cm}{0.8pt}}), $\Gamma_3$ ({\color{green}\rule[0.5ex]{0.5cm}{0.8pt}}), $\Gamma_4$ ({\color{cyan}\rule[0.5ex]{0.5cm}{0.8pt}}), and $\Gamma_5$ ({\color{magenta}\rule[0.5ex]{0.5cm}{0.8pt}}). The locations of $k_c=7.5$ rad m$^{-1}$ are indicated by vertical dashed lines in (a) and (c-e).}
    \label{fig:mss_cygnss}
\end{figure}

Finally, we discuss the implication of the results in surfactant experiments to the MSS anomalies in CYGNSS remote sensing. Figure \ref{fig:mss_cygnss}(a) shows the MSS at sensor 6 as a function of $k_c$ for mechanically generated waves in clean water and with the highest concentration $\Gamma_8$. At CYGNSS cut-off wavenumber $k_c=7.5$ rad m$^{-1}$ indicated in the figure, we see a difference of the two values of MSS which is not significant but larger than the difference in cases of floating particles (figure \ref{fig:mss_kc_micro}). A more quantitative picture is shown in figure \ref{fig:mss_cygnss}(b), which plots the counterpart of figure \ref{fig:avg_growth_rate_wedge_wave}(a) but for $k_c=7.5$ rad m$^{-1}$. Comparing the clean-water result and the result with $\Gamma_8$, we see a reduction of MSS by $O(10\%)$. For wind-wave cases, figure \ref{fig:mss_cygnss}(c)-(e) show the MSS at sensor 6 as a function of $k_c$ for different concentration levels (for which waves are excited). In these figures we see clear difference of MSS even at $k_c=7.5$ rad m$^{-1}$. Quantitatively, with surfactant levels $\Gamma_2$, $\Gamma_3$ and $\Gamma_5$, the MSS is reduced by 8$\%$, 17$\%$ and 50$\%$ relative to the clean-water results for $u_0$=4.29, 6.59 and 9.09 m s$^{-1}$, respectively. Considering the CYGNSS MSS anomaly which is $O(20\%)$ (in terms of reduction from the Katzberg model results), it is clear that the presence of surfactants in wind seas is the most influential factor for this remote sensing application.

\section{Conclusions} \label{sec:conclusion}
In this paper, we investigate the physical mechanisms underlying the MSS anomaly measured by CYGNSS which has been used to track oceanic microplastics. For this purpose, we experimentally study the effect of floating particles and surfactants on surface waves (in terms of the energy and surface roughness) generated by a mechanical wave maker and/or wind. The floating particles are tested with two sizes of 0.5 cm and 0.8 cm, and our results show that their damping effect on surface waves critically depends on the area fraction of coverage, irrespective of the sizes. Damping effects on both energy and MSS are only observed for fractions above $O(5\sim10\%)$, which are much higher than the oceanic microplastics situation of $O(0.1\%)$. For surfactants, experiments with mechanically generated irregular waves show that they generally result in enhanced damping for both the energy and MSS. However, the ``optimal'' concentration level corresponding to the maximum damping for monochromatic waves cannot be identified for irregular waves tested in our study. In the experiments of waves generated by wind, we find that the presence of surfactants significantly suppresses the wave generation, due to its combined effect of reducing the wind stress and increasing the surfactant-induced damping. The wind stress, obtained from the measured wind profile, is further found to follow an exponential dependence on the surface tension, and a power-law dependence on the concentration level of the surfactant (with the power-law exponent linearly related to the wind speed).

The key results of MSS in this paper are also presented for different cut-off wavenumber $k_c$. When considering the CYGNSS cut-off wavenumber $k_c=7.5$ rad m$^{-1}$, we find that the effect of floating particles on MSS is negligible even at the highest fraction above $10\%$ tested in the current work. The mechanically generated waves with surfactants of highest concentration $\Gamma_8$ result in a $O(10\%)$ reduction relative to the MSS in clean water. In comparison, the wind-waves generated with moderate surfactant concentration of $\Gamma_3$ leads to a $O(17\%)$ reduction from the MSS in clean water, for a wind speed about 9 m s$^{-1}$. Considering the MSS anomalies observed by CYGNSS, which corresponds to CYGNSS MSS $O(20\%)$ lower than the results from the standard Katzberg model, we conclude that the effect of surfactants in a wind sea is the most influential factor for this remote sensing application.

\section*{Declaration of Interests}
The authors report no conflict of interest.

\section*{Acknowledgement}
The authors would like to thank Professor James Duncan and An Wang for useful discussions, and Mr. Jason Bundoff, Mr. Alexander Flick and Mr. Jim Smith for technical help at the Marine Hydrodynamics Laboratory. This work was supported in part by NASA Science Mission Directorate contract NNL13AQ00C with the University of Michigan.
\bibliographystyle{jfm}

\bibliography{jfm-instructions}

\begin{thebibliography}{41}
\expandafter\ifx\csname natexlab\endcsname\relax\def\natexlab#1{#1}\fi
\def\au#1{#1} \def\ed#1{#1} \def\yr#1{#1}\def\at#1{#1}\def\jt#1{\textit{#1}}
  \def\bt#1{#1}\def\bvol#1{\textbf{#1}} \def\vol#1{#1} \def\pg#1{#1}
  \def\publ#1{#1}\def\arxiv#1{#1}\def\org#1{#1}\def\st#1{\textit{#1}}

\bibitem[Adak(1998)]{adak1998time}
{\sc \au{Adak, Sudeshna}} \yr{1998}  \at{Time-dependent spectral analysis of
  nonstationary time series}.  \jt{Journal of the American Statistical
  Association}  \bvol{93}~(444),  \pg{1488--1501}.

\bibitem[Barger {\em et~al.\/}(1970)Barger, Garrett, Mollo-Christensen \&
  Ruggles]{Barger1970}
{\sc \au{Barger, W.~R.}, \au{Garrett, W.~D.}, \au{Mollo-Christensen, E.} \&
  \au{Ruggles, K.~W.}} \yr{1970}  \at{Effects of an artificial sea slick upon
  the atmosphere and the ocean}.  \jt{Journal of Applied Meteorology}
  \bvol{9},  \pg{396--400}.

\bibitem[Bock {\em et~al.\/}(1999)Bock, Hara, Frew \&
  McGillis]{Bock1999_experiment}
{\sc \au{Bock, E.~J.}, \au{Hara, T.}, \au{Frew, N.} \& \au{McGillis, W.}}
  \yr{1999}  \at{Relationship between air‐sea gas transfer and short wind
  waves}.  \jt{Journal of Geophysical Research}  \bvol{104},
  \pg{25821--25831}.

\bibitem[Charnock(1955)]{charnock1955wind}
{\sc \au{Charnock, H}} \yr{1955}  \at{Wind stress on a water surface}.
  \jt{Quarterly Journal of the Royal Meteorological Society}  \bvol{81}~(350),
  \pg{639--640}.

\bibitem[Cox \& Munk(1954)]{CoxMunk54}
{\sc \au{Cox, Charles} \& \au{Munk, Walter}} \yr{1954}  \at{Measurement of the
  roughness of the sea surface from photographs of the sun's glitter}.  \jt{J.
  Opt. Soc. Am.}  \bvol{44}~(11),  \pg{838--850}.

\bibitem[C{\'o}zar {\em et~al.\/}(2014)C{\'o}zar, Echevarr{\'\i}a,
  Gonz{\'a}lez-Gordillo, Irigoien, {\'U}beda, Hern{\'a}ndez-Le{\'o}n, Palma,
  Navarro, Garc{\'\i}a-de Lomas, Ruiz, Fern{\'a}ndez-de Puelles \&
  Duarte]{Cozar2014_plastics}
{\sc \au{C{\'o}zar, Andr{\'e}s}, \au{Echevarr{\'\i}a, Fidel},
  \au{Gonz{\'a}lez-Gordillo, J.~Ignacio}, \au{Irigoien, Xabier}, \au{{\'U}beda,
  B{\'a}rbara}, \au{Hern{\'a}ndez-Le{\'o}n, Santiago}, \au{Palma,
  {\'A}lvaro~T.}, \au{Navarro, Sandra}, \au{Garc{\'\i}a-de Lomas, Juan},
  \au{Ruiz, Andrea}, \au{Fern{\'a}ndez-de Puelles, Mar{\'\i}a~L.} \&
  \au{Duarte, Carlos~M.}} \yr{2014}  \at{Plastic debris in the open ocean}.
  \jt{Proceedings of the National Academy of Sciences}  \bvol{111}~(28),
  \pg{10239--10244}.

\bibitem[Ellison(1956)]{ellison1956atmospheric}
{\sc \au{Ellison, T.~H.}} \yr{1956}  \at{Atmospheric turbulence}.  \jt{Surveys
  in mechanics}  \bvol{400},  \pg{430}.

\bibitem[Ermakov {\em et~al.\/}(1986)Ermakov, Zujkova, Panchenko, Salashin,
  Talipova \& Titov]{Ermakov1986}
{\sc \au{Ermakov, S.A.}, \au{Zujkova, A.M.}, \au{Panchenko, A.R.},
  \au{Salashin, S.G.}, \au{Talipova, T.G.} \& \au{Titov, V.I.}} \yr{1986}
  \at{Surface film effect on short wind waves}.  \jt{Dynamics of Atmospheres
  and Oceans}  \bvol{10}~(1),  \pg{31--50}.

\bibitem[Evans \& Ruf(2021)]{EvansRuf2021}
{\sc \au{Evans, Madeline~C.} \& \au{Ruf, Christopher~S.}} \yr{2021}  \at{Toward
  the detection and imaging of ocean microplastics with a spaceborne radar}.
  \jt{IEEE Transactions on Geoscience and Remote Sensing}  \pg{pp. 1--9}.

\bibitem[Hay(1955)]{hay1955some}
{\sc \au{Hay, JS}} \yr{1955}  \at{Some observations of air flow over the sea}.
  \jt{Quarterly Journal of the Royal Meteorological Society}  \bvol{81}~(349),
  \pg{307--319}.

\bibitem[H\"uhnerfuss {\em et~al.\/}(1983)H\"uhnerfuss, Alpers, Garrett, Lange
  \& Stolte]{Huhnerfuss1983_field}
{\sc \au{H\"uhnerfuss, Heinrich}, \au{Alpers, Werner}, \au{Garrett,
  William~D.}, \au{Lange, Philipp~A.} \& \au{Stolte, Siegfried}} \yr{1983}
  \at{Attenuation of capillary and gravity waves at sea by monomolecular
  organic surface films}.  \jt{Journal of Geophysical Research: Oceans}
  \bvol{88}~(C14),  \pg{9809--9816},  \arxiv{arXiv:
  https://agupubs.onlinelibrary.wiley.com/doi/pdf/10.1029/JC088iC14p09809}.

\bibitem[H{\"u}hnerfuss {\em et~al.\/}(1981)H{\"u}hnerfuss, Alpers, Lange \&
  Walter]{huhnerfuss1981attenuation}
{\sc \au{H{\"u}hnerfuss, Heinrich}, \au{Alpers, Werner}, \au{Lange, Philipp~A}
  \& \au{Walter, Wolfgang}} \yr{1981}  \at{Attenuation of wind waves by
  artificial surface films of different chemical structure}.  \jt{Geophysical
  Research Letters}  \bvol{8}~(11),  \pg{1184--1186}.

\bibitem[Katzberg {\em et~al.\/}(2006)Katzberg, Torres \& Ganoe]{katzberg2006}
{\sc \au{Katzberg, Stephen~J}, \au{Torres, Omar} \& \au{Ganoe, George}}
  \yr{2006}  \at{Calibration of reflected gps for tropical storm wind speed
  retrievals}.  \jt{Geophysical Research Letters}  \bvol{33}~(18).

\bibitem[King {\em et~al.\/}(2021)King, Pascual, Clarizia \&
  de~Maagt]{king2021can}
{\sc \au{King, Jennifer}, \au{Pascual, Daniel}, \au{Clarizia, Maria~Paola} \&
  \au{de~Maagt, Peter}} \yr{2021} Can gnss-reflectometry support global
  monitoring of floating matter in the ocean?  \bt{In {\em 2021 IEEE
  International Geoscience and Remote Sensing Symposium IGARSS\/}},  \pg{pp.
  7520--7521}. IEEE.

\bibitem[Lapham {\em et~al.\/}(1999)Lapham, Dowling \& Schultz]{Dowling1999}
{\sc \au{Lapham, Gary~S.}, \au{Dowling, David~R.} \& \au{Schultz, William~W.}}
  \yr{1999}  \at{In situ force-balance tensiometry}.  \jt{Experiments in
  Fluids}  \bvol{27},  \pg{157--166}.

\bibitem[Lapham {\em et~al.\/}(2001)Lapham, Dowling \& Schultz]{Dowling2001}
{\sc \au{Lapham, Gary~S.}, \au{Dowling, David~R.} \& \au{Schultz, William~W.}}
  \yr{2001}  \at{Linear and nonlinear gravity-capillary water waves with a
  soluble surfactant}.  \jt{Experiments in Fluids}  \bvol{30},  \pg{448--457}.

\bibitem[Lin {\em et~al.\/}(2008)Lin, Moeng, Tsai, Sullivan \&
  Belcher]{lin2008direct}
{\sc \au{Lin, Mei-Ying}, \au{Moeng, Chin-Hoh}, \au{Tsai, Wu-Ting},
  \au{Sullivan, Peter~P} \& \au{Belcher, Stephen~E}} \yr{2008}  \at{Direct
  numerical simulation of wind-wave generation processes}.  \jt{Journal of
  Fluid Mechanics}  \bvol{616},  \pg{1--30}.

\bibitem[Liu {\em et~al.\/}(2018)Liu, Garrett, Maddy \&
  Boukabara]{liu2018impact}
{\sc \au{Liu, Ling}, \au{Garrett, Kevin}, \au{Maddy, Eric~S} \& \au{Boukabara,
  Sid-Ahmed}} \yr{2018}  \at{Impact assessment of assimilating nasa’s
  rapidscat surface wind retrievals in the noaa global data assimilation
  system}.  \jt{Monthly Weather Review}  \bvol{146}~(4),  \pg{929--942}.

\bibitem[Liu(2016)]{liu2016wind}
{\sc \au{Liu, Xinan}} \yr{2016}  \at{A laboratory study of spilling breakers in
  the presence of light-wind and surfactants}.  \jt{Journal of Geophysical
  Research: Oceans}  \bvol{121}~(3),  \pg{1846--1865}.

\bibitem[Liu \& Duncan(2006)]{duncan2006JFM}
{\sc \au{Liu, Xinan} \& \au{Duncan, James~H.}} \yr{2006}  \at{An experimental
  study of surfactant effects on spilling breakers}.  \jt{Journal of Fluid
  Mechanics}  \bvol{567},  \pg{433--455}.

\bibitem[Lombardini {\em et~al.\/}(1989)Lombardini, Fiscella, Trivero, Cappa \&
  Garrett]{lombardini1989}
{\sc \au{Lombardini, P.~P.}, \au{Fiscella, B.}, \au{Trivero, P.}, \au{Cappa,
  C.} \& \au{Garrett, W.~D.}} \yr{1989}  \at{Modulation of the spectra of short
  gravity waves by sea surface films: Slick detection and characterization with
  a microwave probe}.  \jt{Journal of Atmospheric and Oceanic Technology}
  \bvol{6}~(6),  \pg{882--890}.

\bibitem[Lucassen(1968)]{lucassen1968longitudinal}
{\sc \au{Lucassen, J}} \yr{1968}  \at{Longitudinal capillary waves. part
  1.—theory}.  \jt{Transactions of the Faraday Society}  \bvol{64},
  \pg{2221--2229}.

\bibitem[Lucassen(1982)]{Lucassen1982EffectOS}
{\sc \au{Lucassen, Jacob}} \yr{1982}  \at{Effect of surface-active material on
  the damping of gravity waves: A reappraisal}.  \jt{Journal of Colloid and
  Interface Science}  \bvol{85},  \pg{52--58}.

\bibitem[Manikantan \& Squires(2020)]{surfactant_hidden_variables}
{\sc \au{Manikantan, Harishankar} \& \au{Squires, Todd~M.}} \yr{2020}
  \at{Surfactant dynamics: hidden variables controlling fluid flows}.
  \jt{Journal of Fluid Mechanics}  \bvol{892},  \pg{P1}.

\bibitem[Mart{\'\i}nez-Vicente {\em et~al.\/}(2019)Mart{\'\i}nez-Vicente,
  Clark, Corradi, Aliani, Arias, Bochow, Bonnery, Cole, C{\'o}zar, Donnelly
  {\em et~al.\/}]{martinez2019measuring}
{\sc \au{Mart{\'\i}nez-Vicente, V{\'\i}ctor}, \au{Clark, James~R}, \au{Corradi,
  Paolo}, \au{Aliani, Stefano}, \au{Arias, Manuel}, \au{Bochow, Mathias},
  \au{Bonnery, Guillaume}, \au{Cole, Matthew}, \au{C{\'o}zar, Andr{\'e}s},
  \au{Donnelly, Rory} \& \au{others}} \yr{2019}  \at{Measuring marine plastic
  debris from space: initial assessment of observation requirements}.
  \jt{Remote Sensing}  \bvol{11}~(20),  \pg{2443}.

\bibitem[Maximenko {\em et~al.\/}(2019)Maximenko, Corradi, Law, Van~Sebille,
  Garaba, Lampitt, Galgani, Martinez-Vicente, Goddijn-Murphy, Veiga {\em
  et~al.\/}]{maximenko2019toward}
{\sc \au{Maximenko, Nikolai}, \au{Corradi, Paolo}, \au{Law, Kara~Lavender},
  \au{Van~Sebille, Erik}, \au{Garaba, Shungudzemwoyo~P}, \au{Lampitt,
  Richard~Stephen}, \au{Galgani, Francois}, \au{Martinez-Vicente, Victor},
  \au{Goddijn-Murphy, Lonneke}, \au{Veiga, Joana~Mira} \& \au{others}}
  \yr{2019}  \at{Toward the integrated marine debris observing system}.
  \jt{Frontiers in marine science}  \bvol{6},  \pg{447}.

\bibitem[Mitsuyasu \& Honda(1986)]{mitsuyasu1986}
{\sc \au{Mitsuyasu, H.} \& \au{Honda, T.}} \yr{1986}  \at{The effects of
  surfactant on certain air—sea interaction phenomena}.  \bt{In {\em Wave
  Dynamics and Radio Probing of the Ocean Surface\/}},  \pg{pp. 95--115}.
  \publ{Springer}.

\bibitem[Phillips(1958)]{phillips_1958}
{\sc \au{Phillips, O.~M.}} \yr{1958}  \at{The equilibrium range in the spectrum
  of wind-generated waves}.  \jt{Journal of Fluid Mechanics}  \bvol{4}~(4),
  \pg{426–434}.

\bibitem[Ruf \& Evans(2020)]{ruf2020detection}
{\sc \au{Ruf, Christopher} \& \au{Evans, Madeline}} \yr{2020} Detection and
  dynamic imaging of ocean microplastics from space.  \bt{In {\em EGU General
  Assembly Conference Abstracts\/}},  \pg{p. 3120}.

\bibitem[Ruf {\em et~al.\/}(2013)Ruf, Unwin, Dickinson, Rose, Rose, Vincent \&
  Lyons]{Ruf2013}
{\sc \au{Ruf, C.}, \au{Unwin, M.}, \au{Dickinson, J.}, \au{Rose, R.}, \au{Rose,
  D.}, \au{Vincent, M.} \& \au{Lyons, A.}} \yr{2013}  \at{Cygnss: Enabling the
  future of hurricane prediction [remote sensing satellites]}.  \jt{IEEE
  Geoscience and Remote Sensing Magazine}  \bvol{1}~(2),  \pg{52--67}.

\bibitem[Ruf {\em et~al.\/}(2016)Ruf, Atlas, Chang, Clarizia, Garrison,
  Gleason, Katzberg, Jelenak, Johnson, Majumdar, O?brien, Posselt, Ridley, Rose
  \& Zavorotny]{Ruf2016}
{\sc \au{Ruf, Christopher~S.}, \au{Atlas, Robert}, \au{Chang, Paul~S.},
  \au{Clarizia, Maria~Paola}, \au{Garrison, James~L.}, \au{Gleason, Scott},
  \au{Katzberg, Stephen~J.}, \au{Jelenak, Zorana}, \au{Johnson, Joel~T.},
  \au{Majumdar, Sharanya~J.}, \au{O?brien, Andrew}, \au{Posselt, Derek~J.},
  \au{Ridley, Aaron~J.}, \au{Rose, Randall~J.} \& \au{Zavorotny, Valery~U.}}
  \yr{2016}  \at{New ocean winds satellite mission to probe hurricanes and
  tropical convection}.  \jt{Bulletin of the American Meteorological Society}
  \bvol{97}~(3),  \pg{385--395}.

\bibitem[Savtchenko {\em et~al.\/}(1997)Savtchenko, Tang \&
  Wu]{Wu1997_circular_tank}
{\sc \au{Savtchenko, A.}, \au{Tang, Shih} \& \au{Wu, Jin}} \yr{1997}
  \at{Effects of surfactant on the growth of wind waves — simultaneous
  observations with optical and microwave sensors}.  \jt{Journal of Marine
  Systems}  \bvol{13},  \pg{273--282}.

\bibitem[van Sebille {\em et~al.\/}(2020)van Sebille, Aliani, Law, Maximenko,
  Alsina, Bagaev, Bergmann, Chapron, Chubarenko, C{\'{o}}zar, Delandmeter,
  Egger, Fox-Kemper, Garaba, Goddijn-Murphy, Hardesty, Hoffman, Isobe,
  Jongedijk, Kaandorp, Khatmullina, Koelmans, Kukulka, Laufkötter, Lebreton,
  Lobelle, Maes, Martinez-Vicente, Maqueda, Poulain-Zarcos, Rodr{\'{\i}}guez,
  Ryan, Shanks, Shim, Suaria, Thiel, van~den Bremer \&
  Wichmann]{van_Sebille_2020}
{\sc \au{van Sebille, Erik}, \au{Aliani, Stefano}, \au{Law, Kara~Lavender},
  \au{Maximenko, Nikolai}, \au{Alsina, Jos{\'{e}}~M}, \au{Bagaev, Andrei},
  \au{Bergmann, Melanie}, \au{Chapron, Bertrand}, \au{Chubarenko, Irina},
  \au{C{\'{o}}zar, Andr{\'{e}}s}, \au{Delandmeter, Philippe}, \au{Egger,
  Matthias}, \au{Fox-Kemper, Baylor}, \au{Garaba, Shungudzemwoyo~P},
  \au{Goddijn-Murphy, Lonneke}, \au{Hardesty, Britta~Denise}, \au{Hoffman,
  Matthew~J}, \au{Isobe, Atsuhiko}, \au{Jongedijk, Cleo~E}, \au{Kaandorp,
  Mikael L~A}, \au{Khatmullina, Liliya}, \au{Koelmans, Albert~A}, \au{Kukulka,
  Tobias}, \au{Laufkötter, Charlotte}, \au{Lebreton, Laurent}, \au{Lobelle,
  Delphine}, \au{Maes, Christophe}, \au{Martinez-Vicente, Victor}, \au{Maqueda,
  Miguel Angel~Morales}, \au{Poulain-Zarcos, Marie}, \au{Rodr{\'{\i}}guez,
  Ernesto}, \au{Ryan, Peter~G}, \au{Shanks, Alan~L}, \au{Shim, Won~Joon},
  \au{Suaria, Giuseppe}, \au{Thiel, Martin}, \au{van~den Bremer, Ton~S} \&
  \au{Wichmann, David}} \yr{2020}  \at{The physical oceanography of the
  transport of floating marine debris}.  \jt{Environmental Research Letters}
  \bvol{15}~(2),  \pg{023003}.

\bibitem[van Sebille {\em et~al.\/}(2015)van Sebille, Wilcox, Lebreton,
  Maximenko, Hardesty, van Franeker, Eriksen, Siegel, Galgani \&
  Law]{van_Sebille_2015}
{\sc \au{van Sebille, Erik}, \au{Wilcox, Chris}, \au{Lebreton, Laurent},
  \au{Maximenko, Nikolai}, \au{Hardesty, Britta~Denise}, \au{van Franeker,
  Jan~A}, \au{Eriksen, Marcus}, \au{Siegel, David}, \au{Galgani, Francois} \&
  \au{Law, Kara~Lavender}} \yr{2015}  \at{A global inventory of small floating
  plastic debris}.  \jt{Environmental Research Letters}  \bvol{10}~(12),
  \pg{124006}.

\bibitem[Stokes(1847)]{stokes_2009}
{\sc \au{Stokes, George~Gabriel}} \yr{1847}  \at{On the theory of oscillatory
  waves}.  \jt{Transactions of the Cambridge Philosophical Society}
  \bvol{8}~(441-55).

\bibitem[Sutherland \& Balmforth(2019)]{Sutherland2019_PRFluids}
{\sc \au{Sutherland, Bruce~R.} \& \au{Balmforth, Neil~J.}} \yr{2019}
  \at{Damping of surface waves by floating particles}.  \jt{Phys. Rev. Fluids}
  \bvol{4},  \pg{014804}.

\bibitem[Tang \& Wu(1992)]{tang1992suppression}
{\sc \au{Tang, Shih} \& \au{Wu, Jin}} \yr{1992}  \at{Suppression of
  wind-generated ripples by natural films: A laboratory study}.  \jt{Journal of
  Geophysical Research: Oceans}  \bvol{97}~(C4),  \pg{5301--5306}.

\bibitem[Uz {\em et~al.\/}(2002)Uz, Donelan, Hara \& Bock]{uz2002_laboratory}
{\sc \au{Uz, B~Mete}, \au{Donelan, Mark~A}, \au{Hara, Tetsu} \& \au{Bock,
  Erik~J}} \yr{2002}  \at{Laboratory studies of wind stress over surface
  waves}.  \jt{Boundary-layer meteorology}  \bvol{102}~(2),  \pg{301--331}.

\bibitem[Wang {\em et~al.\/}(2020)Wang, Zavorotny, Johnson, Yi, Ruf, Gleason,
  McKague, Hwang, Rogers, Chen, Pan \& Bakker]{mss_calibration}
{\sc \au{Wang, Tianlin}, \au{Zavorotny, Valery~U.}, \au{Johnson, Joel}, \au{Yi,
  Yuchan}, \au{Ruf, Christopher}, \au{Gleason, Scott}, \au{McKague, Darren},
  \au{Hwang, Paul}, \au{Rogers, Erick}, \au{Chen, Shuyi}, \au{Pan, Yulin} \&
  \au{Bakker, Thomas}} \yr{2020} Improvement of cygnss level 1 calibration
  using modeling and measurements of ocean surface mean square slope.  \bt{In
  {\em IGARSS 2020 - 2020 IEEE International Geoscience and Remote Sensing
  Symposium\/}},  \pg{pp. 5909--5912}.

\bibitem[Wei \& Wu(1992)]{wei_wu1992}
{\sc \au{Wei, Yi} \& \au{Wu, Jin}} \yr{1992}  \at{In situ measurements of
  surface tension, wave damping, and wind properties modified by natural
  films}.  \jt{Journal of Geophysical Research: Oceans}  \bvol{97}~(C4),
  \pg{5307--5313}.

\bibitem[Wu \& Deike(2021)]{Deike2021_windwave}
{\sc \au{Wu, Jiarong} \& \au{Deike, Luc}} \yr{2021}  \at{Wind wave growth in
  the viscous regime}.  \jt{Phys. Rev. Fluids}  \bvol{6},  \pg{094801}.

\end{thebibliography}

\end{document}